\documentclass{svjour3}                     
\smartqed  
\usepackage[english]{babel}
\usepackage{amsfonts}
\usepackage{amssymb}
\usepackage[all]{xy}
\usepackage{color}
\usepackage[centertags]{amsmath}
\usepackage{latexsym}
\usepackage{graphicx}
\usepackage{textcomp}
\usepackage{tabularx}
\usepackage{t1enc}
\usepackage{multicol}
\usepackage{enumitem}
\usepackage{lineno,hyperref}
\usepackage{cite}

\journalname{Nonlinear Dynamics}
\begin{document}

\title{Nonlinear dynamics of a vertical pendulum driven by magnetic field provided by two coils magnets: analytical, numerical and experimental studies}

\author{B. Nana$^{a*}$, K. Polczy\'{n}ski$^{b}$, P. Woafo$^{c}$ and J. Awrejcewicz$^{b}$.}


\institute{\at
$^{a*}$ Department of Physics, Higher Teacher Training College, University of Bamenda,
PO Box 39 Bamenda, Cameroon. \email{na1bo@yahoo.fr}
\and \at
$^b$ Department of Automation, Biomechanics, and Mechatronics, Lodz University of Technology, 1/15
			Stefanowskiego Str., 90-537 Lodz, Poland
\and \at
$^c$ Laboratory of Modelling and Simulation in Engineering, Biomimetics and Prototypes,
Faculty of Science, University of Yaounde I, PO Box 812 Yaounde, Cameroon.
}

\date{Received: date / Accepted: date}

\maketitle

\begin{abstract}
In the present work, we analyzed theoretically and experimentally the nonlinear dynamics of a magnetic pendulum excited through the interactions of a strong neodymium magnet and two coils placed symmetrically around the zero angular position. The forces between the magnet and coils and generated torques acting on the pendulum are derived using the magnetic charges interaction model and an experimentally fitted model. System equilibrium points are obtained, and their stability is investigated. It is found that when the currents in two coils are negative, the shape of the mechanical potential is bistable. The bistable potential might be symmetric if the currents have the same values and asymmetric when they are different. Asymmetric bistable potential is observed when coil currents have different signs. However, in the case of positive coil currents, a symmetric tristable potential is detected when the currents are the same, and an asymmetric tristable potential takes place when the positive currents have different values. Considering the sinusoidal coil current signals, analytical calculations using the harmonic balance method and numerical simulations are carried out for this electric-magneto-mechanical system. The obtained results are shown in terms of frequency-response diagrams, displacement time series, and phase portraits. The two-parameter bifurcation diagrams are plotted showing the different dynamical behaviors considering the current amplitudes and frequency as the control parameters. Amplitude jumps, hysteresis, and multistability are also observed. Some phase portraits and the coexistence of attractors are obtained numerically and confirmed experimentally. A good agreement between the numerical simulation and experimental measurement is achieved.
\end{abstract}


\section{Introduction}

$\quad$ In recent years, several theoretical and experimental studies demonstrating chaotic motion in pendulum systems have been published. This has motivated intense research that has led to the discovery of many novel features of pendulum systems \cite{bb1,bb2,bb3,bb4,bb5,bb6,bb7,bb8,bb9}. Mechanical systems including the pendulum arise in many practical applications in mechanical/mechatronic engineering such as machines and mechanisms, control and automation devices, space exploration machines, sensors, robots, and servos construction, as well as several MEMS and NEMS devices \cite{bb10,bb11,bb12,bb13,bb14,bb15,bb16,bb17,bb18,bb19,bb20,bb21,bb22}.

Recently, there has been increasing interest in the design and implementation of magnetic pendulums \cite{bb23,bb24,bb25,bb26,bb27,bb28,bb29,bb30,bb31}. A magnetic pendulum is considered as a mechanical system in which the upper end is fixed to a rotational axis, and a permanent magnet or ferromagnetic bob is installed on its other end. The unit composed of a pendulum-magnet moves under the action of another permanent magnet or electromagnet located close to its lower end. In general, the vertical magnetic pendulum is affected by two potential energies; namely, the gravitational potential energy, and the magnetic potential energy.

An interesting system based on a magnetic pendulum for positioning offshore wind turbines during their installation has been presented by Atzampou et al. \cite{Atzampou2024}. The contactless motion compensation technique has been validated numerically and experimentally. The obtained results have illuminated the path for further potential to increase the efficiency of offshore wind installations.

Pilipchuk et al. \cite{PilipchukPolczynski2022} have studied a system composed of two weakly coupled magnetic pendulums. The authors have developed algorithms for controlling an energy transfer between pendulums using a proper magnetic field generated by coils placed under the pendulums. Numerical, analytical, and experimental investigations have been performed which have proved that analyzed transfer energy control algorithms work properly and can be used in energy harvesting and vibration-damping systems.

Magnetic pendulum arrays were recently employed as an antenna for efficient transmission in ultra-low frequencies (ULF) \cite{Fereidoony2022a, SrinivasPrasad2020}. The antenna was composed of 28 cylindrical magnets. Assuming that magnets behave like magnetic pendulums, the authors analyzed their vibrations and efficiency in generating electromagnetic waves in the ultra-low frequency range. As a result, it turned out that the efficiency of this type of antenna is 7 dB higher than that of an ordinary electric antenna.

Inspired by the results of papers \cite{NanaWoafoPolczynski_2024,bb28} regarding a magnetic pendulum subjected to the action of one coil at its base and those reported in Ref. \cite{bb29} for horizontal magnetic pendulum subjected to the action of fixed magnets placed symmetrically to the reference horizontal direction, this work is aimed to analyze the dynamical behavior of a magnetic pendulum driven by two coils.\textcolor[rgb]{1.00,0.00,0.00}{ The permanent magnet is positioned at the lower end of the pendulum and the axis of the coils are symmetric with respect to the pendulum vertical position. Compared to the other works mentioned previously, this system presents a bistable or tristable potential depending on the parameters. Hence, the objective of this work is to analyze the effects of such potential on the dynamical behavior of the system. The analysis employs mathematical, numerical, and experimental methods.} In this work, magnet and coils are considered as magnetic charges and we have used an equivalent Coulomb law to derive the formulas of the magnetic torques governing coil-magnet pair interaction \cite{bb25,bb26,Grzelczyk2024}. This magnetic charge interaction model has been used recently by the authors of reference \cite{bb25} and excellent agreement was found between their theoretical and experimental results.

The paper is structured in the following order. The physical description and the mathematical model of the system are presented in Section 2. The equilibrium points and their stability as well as the shapes of the system and the shapes of the corresponding potential are analyzed in Section 3 where the solenoids are powered with constant current sources. In Section 4, the current is taken to be a sinusoidal one. We find mathematically the frequency response of the system invoking the harmonic balance method. Then, we apply the numerical simulation to complement the mathematical results and plot the numerous bifurcation diagrams exhibiting novel dynamic phenomena. In Section 5, comparisons between our simulations and experimental results are presented. Our conclusions are contained in the final section and present concluding remarks of our research.

\section{Physical and mathematical models}

\subsection{Physical model}

\begin{figure}[h!]
\begin{center}
\includegraphics[width=7.5cm,height=6.5cm]{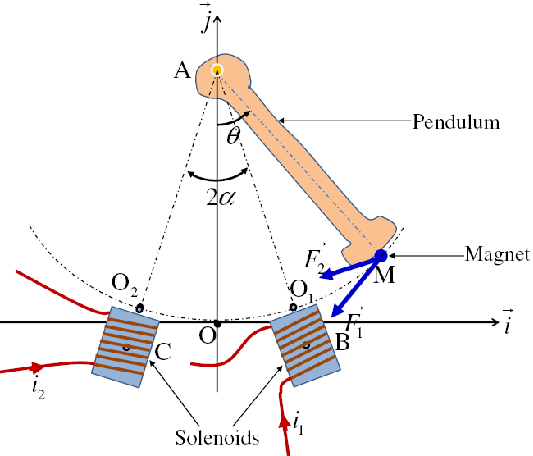}\hspace*{0cm}
\caption{\label{FIGURE1}\footnotesize {Pendulum-magnet embedded into a variable magnetic field invoked by two current-carrying solenoids.}}
\end{center}
\end{figure}

$\quad$ A schematic representation of the magnetic pendulum driven by two magnetic forces is shown in Figure \ref{FIGURE1}, where $\theta$ is the angular deviation.

It is assumed that the pendulum motion is confined to the $(x,\,y)$-plane and that the pivot is at the origin of the coordinate system. The potential energy of the pendulum is caused by the gravity field. The level of zero potential is considered to be the $xz$ plane of the chosen coordinate system. As indicated in the figure, a strong neodymium magnet is joined to the free end of the pendulum while the upper end of the pendulum is fixed to a horizontal rotational axis. Two current-carrying solenoids are placed under the pendulum arm, and the axis of the first solenoid located at point B is attributed to an angle $\alpha$ with the vertical axis, namely the $y-$axis. Similarly, the axis of the second solenoid located at point C makes an angle $-\alpha$ with the vertical axis. Points B and C are the centers of the first and second solenoids, respectively. Let $\overline{O_1B}$=$\overline{O_2C}$=r and $\ell$ represents the length of the pendulum. Since the magnet is fixed on the rod, its angular velocity is equal to the angular velocity of the pendulum.

\subsection{Mathematical modeling}

$\quad$ In this study, to simplify our analysis, the neodymium magnet and the solenoids are considered to be point magnets. \textcolor[rgb]{1.00,0.00,0.00}{The magnetic forces $\overrightarrow{F_1}$ and $\overrightarrow{F_2}$ produced, respectively, by the first and second solenoids on the neodymium magnet have the following forms \cite{bb25}
\begin{equation}\label{EQUAT1}
\overrightarrow {F_1 }  = \frac{{\mu _0 Q_1 Q}}{{4\pi }}\frac{{\overrightarrow {BM} }}{{\left\| {\overrightarrow {BM} } \right\|^3 }} = \frac{{\mu _0 Q_1 Q}}{{4\pi }}\frac{{\left( {\ell \sin \theta  - \left( {\ell  + r} \right)\sin \alpha } \right)\overrightarrow i  - \left( {\ell \cos \theta  - \left( {\ell  + r} \right)\cos \alpha } \right)\overrightarrow j }}{{\left( {2\ell \left( {\ell  + r} \right)\left( {1 - \cos \left( {\theta  - \alpha } \right)} \right) + r^2 } \right)^{\frac{3}{2}} }},
\end{equation}
\begin{equation}\label{EQUAT2}
\overrightarrow {F_2 }  = \frac{{\mu _0 Q_2 Q}}{{4\pi }}\frac{{\overrightarrow {CM} }}{{\left\| {\overrightarrow {CM} } \right\|^3 }} =  \frac{{\mu _0 Q_2 Q}}{{4\pi }}\frac{{\left( {\ell \sin \theta  + \left( {\ell  + r} \right)\sin \alpha } \right)\overrightarrow i  - \left( {\ell \cos \theta  - \left( {\ell  + r} \right)\cos \alpha } \right)\overrightarrow j }}{{\left( {2\ell \left( {\ell  + r} \right)\left( {1 - \cos \left( {\theta  + \alpha } \right)} \right) + r^2 } \right)^{\frac{3}{2}} }}.
\end{equation}}

Here, $\mu_0$ is the permeability of the vacuum, $Q_1$, $Q_2$ and $Q$ are the pole strengths of the solenoids and neodymium magnet, respectively. Since the magnetic fields provided by the solenoids are governed by the currents $i_1(t)$ and $i_2(t)$ flowing through the respective windings, the strengths $Q_1$ and $Q_2$ are respectively proportional to the currents $i_1(t)$ and $i_2(t)$. They can be recast to the following forms: $Q_1=ki_1(t)$ and $Q_2=ki_2(t)$, where $k$ stands for a constant parameter. The magnetic torques provided by the above magnetic forces $\overrightarrow{F_1}$ and $\overrightarrow{F_2}$ follow
\begin{equation}\label{EQUAT3}
\Gamma _a  = \frac{{\mu _0 ki_1 Q}}{{4\pi }}\frac{{\ell \left( {\ell  + r} \right)\sin \left( {\theta  - \alpha } \right)}}{{\left( {2\ell \left( {\ell  + r} \right)\left( {1 - \cos \left( {\theta  - \alpha } \right)} \right) + r^2 } \right)^{\frac{3}{2}} }},
\end{equation}
\begin{equation}\label{EQUAT4}
\Gamma _b  = \frac{{\mu _0 ki_2 Q}}{{4\pi }}\frac{{\ell \left( {\ell  + r} \right)\sin \left( {\theta  + \alpha } \right)}}{{\left( {2\ell \left( {\ell  + r} \right)\left( {1 - \cos \left( {\theta  + \alpha } \right)} \right) + r^2 } \right)^{\frac{3}{2}} }}.
\end{equation}
	After some algebraic transformations, the final magnetic torques take the following final form
\begin{equation}\label{EQUAT5}
\begin{gathered}
\Gamma_1= \frac{{\sigma i_1 \sin \left( {\theta  - \alpha } \right)}}{{\left( {1 + \varepsilon  - \cos \left( {\theta  - \alpha } \right)} \right)^{\frac{3}{2}} }} + \frac{{\sigma i_2 \sin \left( {\theta  + \alpha } \right)}}
{{\left( {1 + \varepsilon  - \cos \left( {\theta  + \alpha } \right)} \right)^{\frac{3}{2}} }},\; \hfill \\
  {\text{where}}\;\;\sigma  = \frac{{k\mu _0 Q}}{{8\pi \sqrt {2\ell \left( {\ell  + r} \right)} }}\;\;{\text{and}}\;\;\varepsilon  = \frac{{r^2 }}{{2\ell \left( {\ell  + r} \right)}}. \hfill \\
\end{gathered}
\end{equation}
Since the currents $i_1$ and $i_2$ are provided by current sources, the effects of the induced electromotive forces attributed to the movement of the neodymium magnet in the solenoids are neglected. The application of Newton's second law to the pendulum yields the following equation of motion
\begin{equation}\label{EQUAT6}
J\ddot \theta  + \beta \dot \theta + C\theta  + mg\ell_1 \sin \theta+ \Gamma _s \left( {\dot \theta } \right)  = \Gamma_1.
\end{equation}
Terms $J$ and $m$ are the pendulum moment of inertia and mass of the pendulum with the fixed magnet, respectively. Term $\ell_1$ represents the distance from the pivot to the center of mass of the magnetic pendulum, $\beta$ is a complete viscous damping coefficient, $C$ is the stiffness of the joint and the parameter $g$ stands for the gravitational acceleration. Finally, $\Gamma _s \left( {\dot \theta } \right)$ represents the nonlinear part of the Stribeck friction torque and is defined as \cite{bb26,bb28}
\begin{equation}\label{EQUAT8}
\Gamma _s \left( {\dot \theta } \right) = \left[ {M_c + \left( {M_s - M_c } \right)\exp \left( { - \frac{{\dot \theta ^2 }}{{v_s^2 }}} \right)} \right]\tanh \left( {\chi \dot \theta } \right).
\end{equation}
In the above equation, $M_c$ represents the magnitude of Coulomb friction, while $M_s$ is the static friction value. The parameter $v_s$ stands for Stribeck velocity, while $\chi$ is a regularization parameter \cite{bb26,bb28}.

Based on the experimental measurements carried out by the authors of reference \cite{bb28}, the system can also be described by the following second-order differential equation:
\begin{equation}\label{EQUAT6A}
J\ddot \theta  + \beta \dot \theta  + C\theta  + mg\ell_1 \sin \theta  + \Gamma _s \left( {\dot \theta } \right) = \Gamma _2,
\end{equation}
where $\Gamma_2$ is the torque exerted by the coils on the neodymium magnet. As defined in equation (\ref{EQUAT5}), $\Gamma_2$ is a function of the coil currents $i_1$ and $i_2$ and is related to the mechanical displacement $\theta$:
\begin{equation}\label{EQUAT5A}
\Gamma _2  = \frac{{2i_1 a\left( {\theta  - \alpha } \right)}}{b}\exp \left( { - \frac{{\left( {\theta  - \alpha } \right)^2 }}{b}} \right) + \frac{{2i_2 a\left( {\theta  + \alpha } \right)}}
{b}\exp \left( { - \frac{{\left( {\theta  + \alpha } \right)^2 }}{b}} \right).
\end{equation}
The fitting parameters $a$ and $b$ are also functions of the amplitudes of the currents through the coil. In what follows, we will assume that there are constants. The parameters used in this work follow the experimental results presented in reference \cite{bb28} and are given in Table \ref{CH1T1}.\\
\begin{table}[h]
\begin{center}
\begin{tabular}{cl cl}
\hline
Parameters	&  Values &	Parameters	&  Values\\
\hline
$a$ & $4.253 \cdot 10^{ - 2} \;{\text{Nm rad}}$ & $b$ & $1.818 \cdot 10^{ - 2} \;{\text{rad}^{\text{2}}}$\\
$r$ & $0.100$ m & $\ell$ & $1.200$ m\\
$J$ & $6.787 \cdot 10^{ - 4} \;{\text{kgm}}^{\text{2}}$ & $\gamma  = mg\ell _1$ & $5.800 \cdot 10^{ - 2} \;{\text{Nm}}$\\
$\beta$ & $2.019 \cdot 10^{ - 4} \;{\text{Nm/rad}}$ & $C$ & $1.742 \cdot 10^{ - 2} \;{\text{Nm/rad}}$\\
$M_c$ & $2.223 \cdot 10^{ - 4} \;{\text{Nm}}$ & $M_s$ & $4.436 \cdot 10^{ - 4} \;{\text{Nm}}$\\
$v_s$ & $5.374 \cdot 10^{ - 1} \;{\text{rad/s}}$ & $\chi$ & $5.759 $\\
$f$ & $0.2{\text{ Hz}} \leqslant f \leqslant 10.0{\text{ Hz}}$ & $I_{m1},\,I_{m2}$ & $0.0{\text{ A}} \leqslant I_{m1}=I_{m2}  \leqslant 5{\text{ A}}$ \\
$I_{1},\,I_{2}$ & $0.0{\text{ A}} \leqslant I_{1}=I_{2} \leqslant 5{\text{ A}}$ &  &
\\\hline
\end{tabular}
\caption{Values of the parameters used in this work \cite{bb28}.}\label{CH1T1}
\end{center}
\end{table}

\section{States of the system powered by constant currents}

\subsection{Equilibrium points}

$\quad$ Firstly, we determined and analyzed the structure of the equilibrium points of the system. In this regard, the system is not subjected to variable excitation, and therefore, the variable current sources $i_1$ and $i_2$ are replaced by constant current sources $I_1$ and $I_2$, respectively. With constant current sources, a solution to equation (\ref{EQUAT6}) is expected to approach an equilibrium point. Setting the derivatives to zero in equation (\ref{EQUAT6}) yields
\begin{equation}\label{EQUAT10}
\frac{{\sigma I_1 \sin \left( {\theta  - \alpha } \right)}}{{\left( {1 + \varepsilon  - \cos \left( {\theta  - \alpha } \right)} \right)^{\frac{3}{2}} }} + \frac{{\sigma I_2 \sin \left( {\theta  + \alpha } \right)}}{{\left( {1 + \varepsilon  - \cos \left( {\theta  + \alpha } \right)} \right)^{\frac{3}{2}} }} - C\theta  - \gamma \sin \theta  = 0.
\end{equation}
After some algebra manipulations, the approximated solutions of equation (\ref{EQUAT10}) are roots of the following polynomial
\begin{equation}\label{EQUAT11}
\left( {D_1  + E_1 \theta } \right)\prod\limits_{n = 1}^2 {\left[ {A_{1n}  + 2B_{1n} \left( {\theta  + \left( { - 1} \right)^n \alpha } \right)^2  + C_{1n} \left( {\theta  + \left( { - 1} \right)^n \alpha } \right)^4 } \right]}  = 0,
\end{equation}
where
\begin{equation}\label{EQUAT12}
\begin{gathered}
  D_1  = \sigma \left( {I_1  - I_2 } \right)\left( {1 + \varepsilon  - \cos \alpha } \right)\sin \alpha , \hfill \\
  E_1  = \frac{\sigma }{2}\left( {3 - 2\cos \alpha  - 2\varepsilon \cos \alpha  - \cos ^2 \alpha } \right)\left( {I_1  + I_2 } \right) + \left( {C + \gamma } \right)\left( {1 + \varepsilon  - \cos \alpha } \right)^{\frac{5}
{2}} , \hfill \\  A_{1n}  = 480\left( {C\varepsilon ^2 \cos \alpha  + \gamma \varepsilon ^2  - 2\sigma I_n \sqrt \varepsilon  \cos \alpha } \right), \hfill \\  B_{1n}  = 20\left( {9C\varepsilon \cos \alpha  + 9\gamma \varepsilon  - 2\gamma \varepsilon ^2  + 4\sigma I_n \sqrt \varepsilon  \cos \alpha } \right), \hfill \\  C_{1n}  = 45C\cos \alpha  - 30C\varepsilon \cos \alpha  - 90\gamma \varepsilon  + 45\gamma  + 4\gamma \varepsilon ^2  - 8\sigma I_n \sqrt \varepsilon  \cos \alpha . \hfill \\\end{gathered}
\end{equation}
Coefficients $D_1$ and $E_1$ are the coefficients of the first order series expansion of equation (\ref{EQUAT10}) around $\theta=0$ rad. Similarly, the coefficients $A_{1n}$, $2B_{1n}$ and  $C_{1n}$ with $n=1,\,2$ are obtained by finding the fourth order series expansion of equation (\ref{EQUAT10}) around the points $\theta=\alpha$ and $\theta=-\alpha$, respectively.
It is found that the possible real roots of equation (\ref{EQUAT11}) have the following forms:
\begin{equation}\label{EQUAT13}
\begin{gathered}
  \theta _{a}  =  - \alpha  - \sqrt {\frac{{ - B_{12}  + \sqrt {B_{12}^2  - A_{12} C_{12} } }}{{C_{12} }}} ,\;\theta _{b}  =  - \alpha ,\;\theta _{o}  =  - \frac{{D_1 }}{{E_1 }},\;\theta _{c}  = \alpha {\text{ and }} \hfill \\
  \theta _{d}  = \alpha  + \sqrt {\frac{{ - B_{11} + \sqrt {B_{11}^2 - A_{11} C_{11} } }}
{{C_{11} }}}, \hfill \\\end{gathered}
\end{equation}

\begin{figure}[t]
\begin{center}
\includegraphics[width=6.5cm,height=4.6cm]{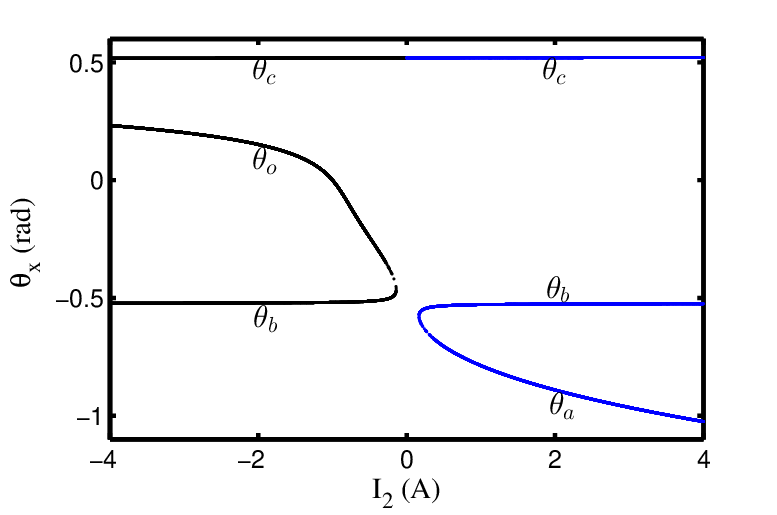}\hspace*{0.0cm}
\includegraphics[width=6.5cm,height=4.6cm]{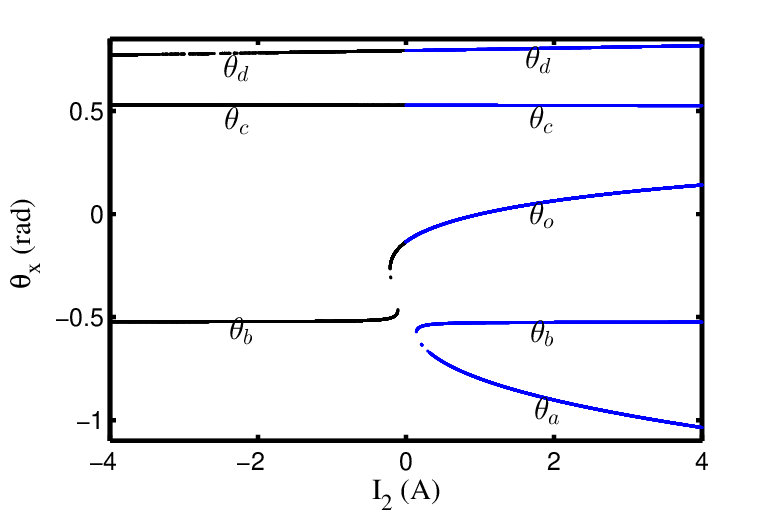}\\
\hspace*{0.50cm} a) \hspace*{6.5cm} b) \hfill\\
\caption{\label{FIGURE2}\footnotesize {Location of the equilibrium points obtained for $\alpha=\frac{\pi}{6}$ rad, and a) for $I_1=-1.0$ A, while b) for $I_1=1.0$ A.}}
\end{center}
\end{figure}

To carry out the analysis in a general way, we consider that the currents $I_1$ and $I_2$ can independently have positive and negative values. Assuming the positive values of $Q$, the number of equilibrium points is defined through the signs of the applied currents $I_1$ and $I_2$, and four important cases are obtained. Figure \ref{FIGURE2} compares the approximate solutions (\ref{EQUAT13}) to the ones obtained using the Newton-Raphson method to solve the equation (\ref{EQUAT10}) considering $I_2$ as the control parameter after fixing $I_1$.

  \begin{enumerate}[label=(\roman*)]
	\item $I_1\leqslant 0$ and $I_2\leqslant 0$; there are three equilibrium points of the system, namely $\theta _{b}$, $\theta _{o}$ and $\theta _{c}$. This situation is illustrated in Figure \ref{FIGURE2}a where the roots of equation (\ref{EQUAT10}) are plotted in black using the Newton-Raphson algorithm and for $I_1=-1.0$ A.
	\item  $I_1\leqslant 0$ and $I_2>0$; also three equilibrium points are obtained here. There are $\theta _{a}$, $\theta _{b}$ and $\theta _{c}$ as shown in the graph of Figure \ref{FIGURE2}a with the curves plotted in blue and for $I_1=-1.0$ A.
	\item $I_1>0$ and $I_2\leqslant 0$; the model still presents three equilibrium points given as $\theta _{b}$, $\theta _{c}$ and $\theta _{d}$. For verification, the roots of equations (\ref{EQUAT11}) are plotted with black in Figure \ref{FIGURE2}b for $I_1=1.0$ A.
	\item $I_1>0$ and $I_2>0$; in this configuration, there are the following five equilibrium points: $\theta _{a}$, $\theta _{b}$, $\theta _{o}$, $\theta _{c}$ and $\theta _{d}$. This situation is highlighted with the curve in blue presented in Figure \ref{FIGURE2}b where the roots of equations (\ref{EQUAT10}) are obtained using the Newton-Raphson algorithm for $I_1=1.0$ A.
\end{enumerate}

Notably, Figure \ref{FIGURE2} presents excellent agreement between the approximate and numerical mathematical solutions.

\subsection{Stability of the equilibrium points}

$\quad$ According to the first, second, and third cases mentioned above, the first and third points are stable while the second point is unstable. The behavior is different in the last situation where the first, third, and fifth equilibrium points are stable while the second and the fourth are unstable. The stability of each equilibrium point $\theta_x$ is estimated based on the eigenvalues $\lambda$ of the corresponding Jacobian matrix
\begin{equation}\label{EQUAT13a}
M_J  = \left[ {\begin{array}{*{20}c}
0 & 1  \\   {\Gamma _e \left( {\theta _x } \right)} & { - \frac{{\beta  + \chi M_s }}
{J}}  \\ \end{array} } \right],
\end{equation}
where
\begin{equation}\label{EQUAT13b}
\begin{gathered}
\Gamma _e \left( \theta  \right) = \frac{{\sigma i_1 }}{{4J}}\frac{{\cos \left( {2\theta  - 2\alpha } \right) + 4\left( {1 + \varepsilon } \right)\cos \left( {\theta  - \alpha } \right) - 5}}{{\left[ {1 + \varepsilon  - \cos \left( {\theta  - \alpha } \right)} \right]^{\frac{5}{2}} }} + \frac{{\sigma i_2 }}{{4J}}\frac{{\cos \left( {2\theta  + 2\alpha } \right) + 4\left( {1 + \varepsilon } \right)\cos \left( {\theta  + \alpha } \right) - 5}}
{{\left[ {1 + \varepsilon  - \cos \left( {\theta  + \alpha } \right)} \right]^{\frac{5}
{2}} }} -  \hfill \\  \qquad \qquad \frac{{C + mg\ell _1 \cos \left( \theta  \right)}}{J}. \hfill \\
\end{gathered}
\end{equation}

\begin{figure}[t]
\begin{center}
\includegraphics[width=6.5cm,height=4.70cm]{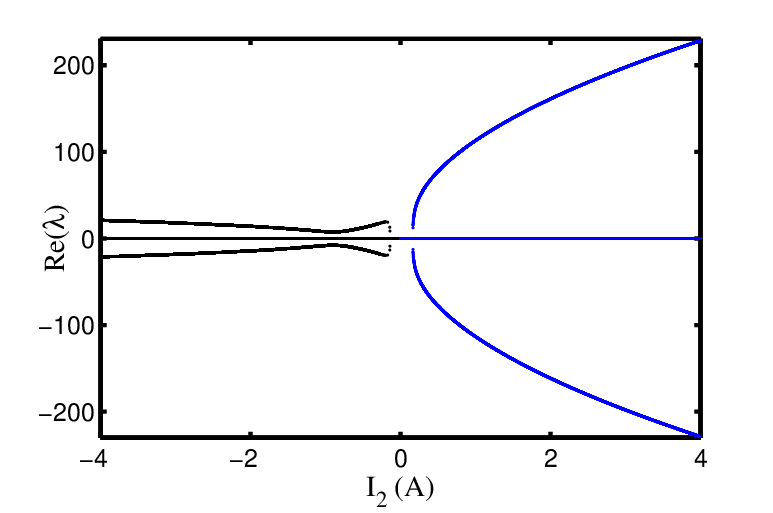}\hspace*{0.0cm}
\includegraphics[width=6.5cm,height=4.70cm]{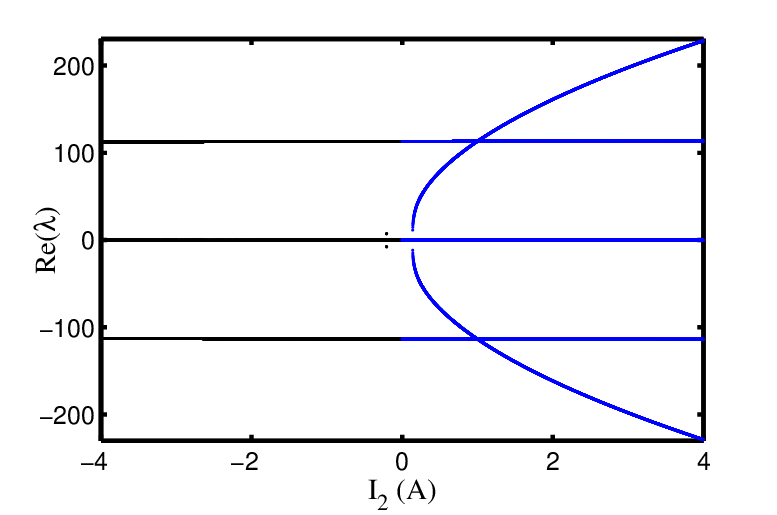}\\
\hspace*{0.5cm} a) \hspace*{6.5cm} b) \hfill\\
\caption{\label{FIGURE3}\footnotesize {Real parts of the eigenvalues of the corresponding Jacobian matrix obtained for $\alpha=\frac{\pi}{6}$ rad and a) for $I_1=-1.5$ A, b) for $I_1=1.5$ A.}}
\end{center}
\end{figure}

The real parts of the eigenvalues $\lambda$ are reported in Figures \ref{FIGURE3}a and \ref{FIGURE3}b plotted for $I_1=-1.0$ A and $I_1=1.0$ A, respectively.

The current $I_2$ is used as the control parameter and $\alpha=\frac{\pi}{6}$. Figure \ref{FIGURE3}a shows two stable and one unstable equilibrium points. On the other hand, Figure \ref{FIGURE3}b presents two stable and one unstable equilibrium points when $I_2$ is negative. For positive $I_2$,~ there are three stable and two unstable equilibrium points. The black and blue colors imply the locations of the equilibrium points presented previously.

\subsection{Shapes of the potential}

$\quad$ The potential $U$ associated with the differential equation (\ref{EQUAT6}) has the following form
\begin{equation}\label{EQUAT14}
U= \frac{{2\sigma I_1 }}{{\sqrt {1 + \varepsilon - \cos \left( {\theta  - \alpha } \right)} }} + \frac{{2\sigma I_2 }}{{\sqrt {1 + \varepsilon - \cos \left( {\theta  + \alpha } \right)} }} + \frac{1}
{2}C\theta ^2 - \gamma \cos \theta.
\end{equation}

\begin{figure}[h!]
\begin{center}
\includegraphics[width=6.5cm,height=4.0cm]{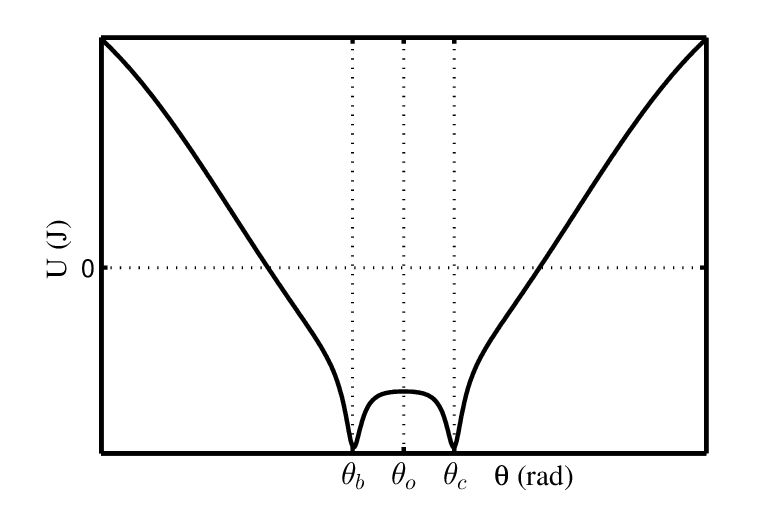}\hspace*{0.0cm}
\includegraphics[width=6.5cm,height=4.0cm]{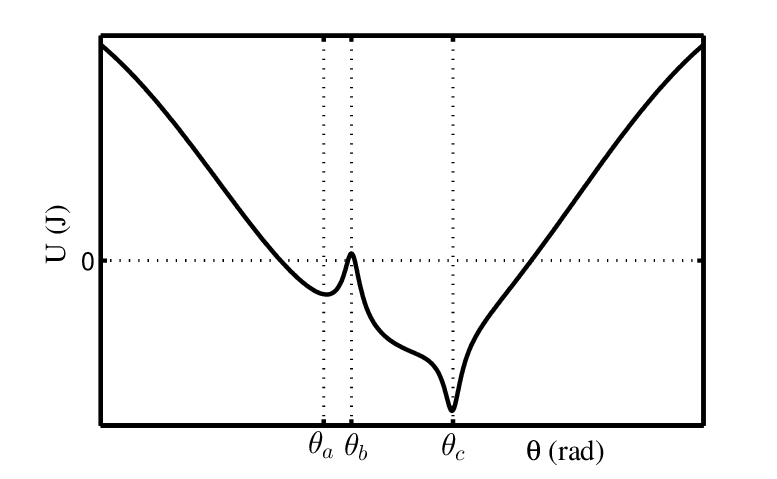}\\
\hspace*{1.0cm} a) \hspace*{6.5cm} b) \hfill\\
\includegraphics[width=6.5cm,height=4.0cm]{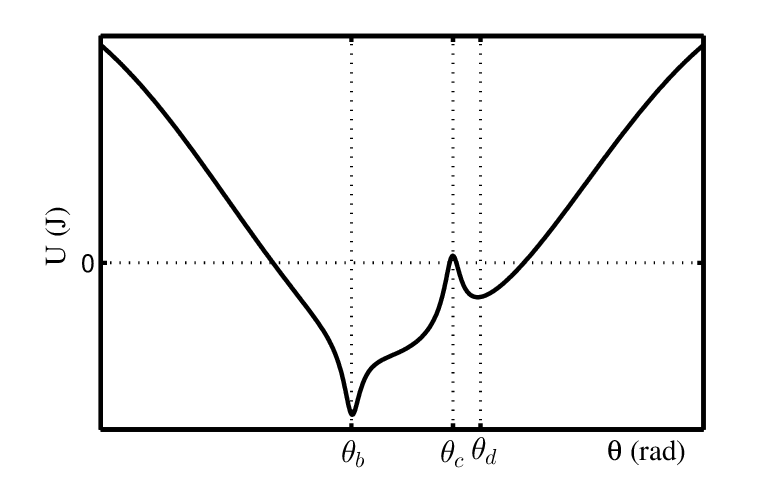}\hspace*{0.0cm}
\includegraphics[width=6.5cm,height=4.0cm]{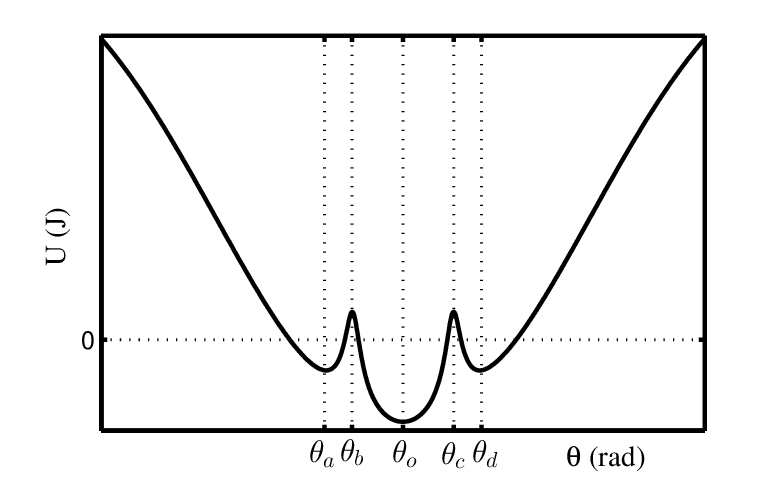}\\
\hspace*{0.50cm} c) \hspace*{6.5cm} d) \hfill\\
\caption{\label{FIGURE4}\footnotesize {Shape of the potential $U$ for different values of the coil currents $I_1$, $I_2$ and for $\alpha=\frac{\pi}{6}$ rad. a) $I_1=I_2=-1.0$ A. b) $I_1=-I_2=-1.0$ A. c) $I_1=-I_2=1.0$ A. d) $I_1=I_2=1.0$ A.}}
\end{center}
\end{figure}

Different shapes of the potential are presented in Figure \ref{FIGURE4} for some configurations of the currents $I_1$ and $I_2$. As demonstrated before in stability analysis of equilibrium points, Figure \ref{FIGURE4} presents globally three kinds of potential depending on the signs of the currents. The first one in Figure \ref{FIGURE4}a corresponds to a bistable shape (two currents are equal and negative). The second form stands for the asymmetric bistable potential (currents have different signs). The orientation of the asymmetry depends on which current is negative or positive as can be seen in Figures \ref{FIGURE4}b and \ref{FIGURE4}c. The last tristable form is presented in Figure \ref{FIGURE4}d, and it takes place when the two currents are positive and equal. Observed that the symmetric bistable shape from Figure \ref{FIGURE4}a and symmetric tristable shape can be deformed to asymmetric bistable or asymmetric tristable shapes if the currents have the same signs, but are different.

\begin{figure}[h!]
\begin{center}
\includegraphics[width=6.5cm,height=4.75cm]{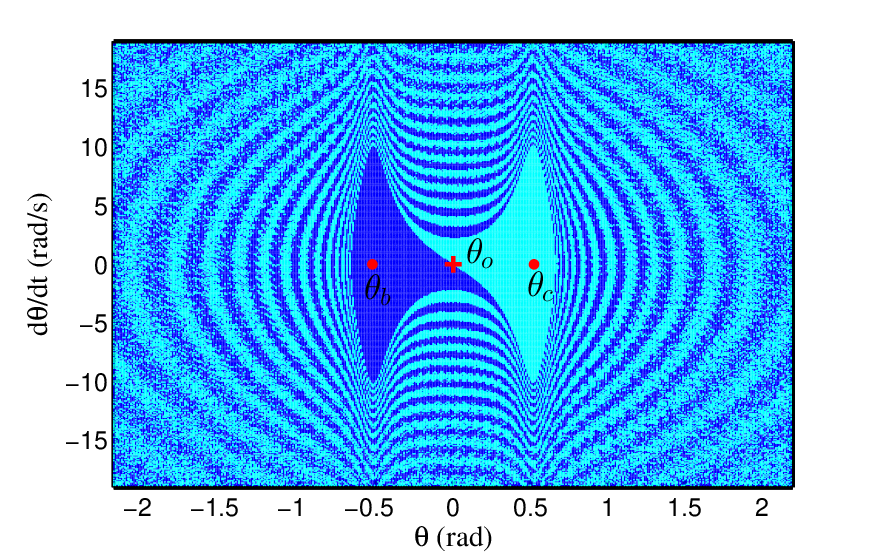}\hspace*{0.0cm}
\includegraphics[width=6.5cm,height=4.75cm]{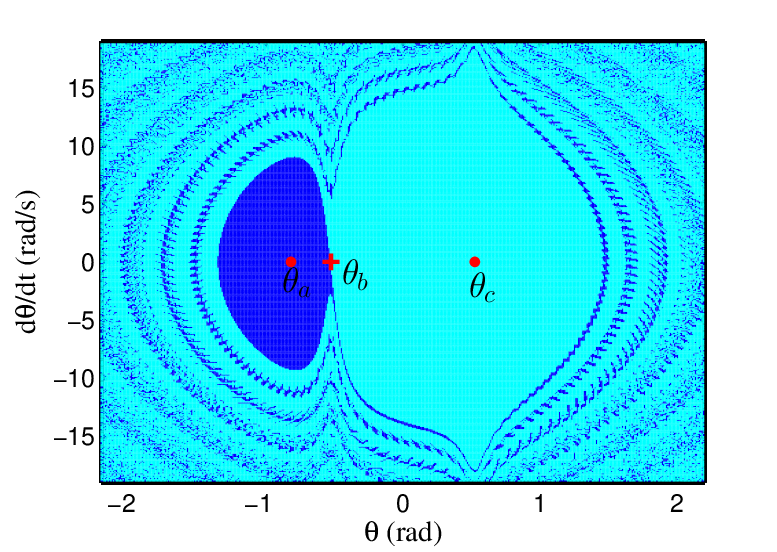}\\
\hspace*{0.5cm} a) \hspace*{6.5cm} b) \hfill\\
\includegraphics[width=6.5cm,height=4.75cm]{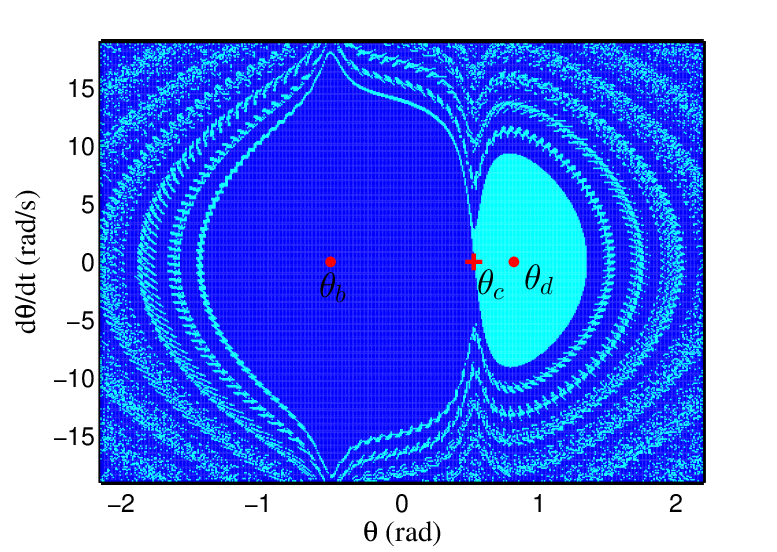}\hspace*{0.0cm}
\includegraphics[width=6.5cm,height=4.75cm]{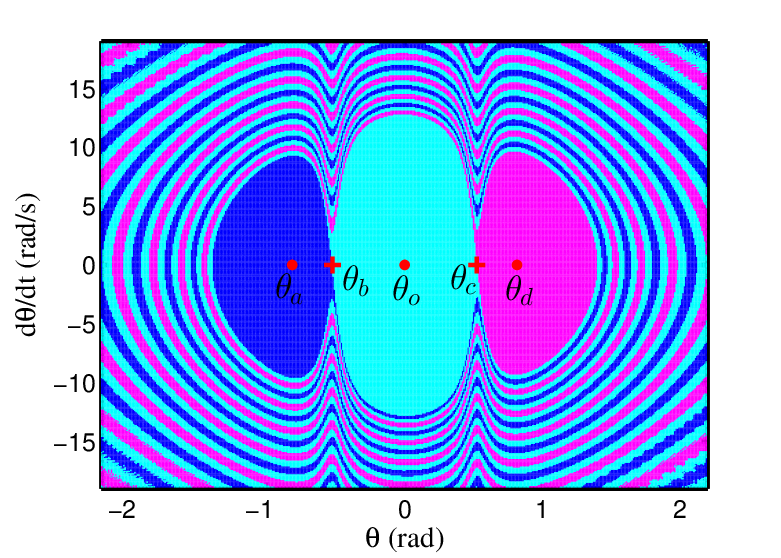}\\
\hspace*{0.5cm} c) \hspace*{6.5cm} d) \hfill\\
\caption{\label{FIGURE5}\footnotesize {Basins of attraction of the stationary solutions for different values of the currents $I_1$ and $I_2$ and for $\alpha=\frac{\pi}{6}$ rad: a) $I_1=I_2=-1.0$ A; b) $I_1=-I_2=-1.0$ A; c) $I_1=-I_2=1.0$ A; d) $I_1=I_2=1.0$ A.}}
\end{center}
\end{figure}

Figure \ref{FIGURE5} presents the basin of attractions for the stationary solutions with regard to the initial conditions of the pendulum, i.e. its angular displacement and angular velocity.
The red points and red crosses indicate the stable and unstable equilibrium points, respectively.
\textcolor[rgb]{1.00,0.00,0.00}{Any initially condition inside the blue regions of Figures \ref{FIGURE5}a, \ref{FIGURE5}b, \ref{FIGURE5}c) and \ref{FIGURE5}d) will ultimately arrive at the stationary solution $\theta_b$, $\theta_a$, $\theta_b$ and $\theta_a$, respectively. For initial conditions inside the cyan regions of Figures \ref{FIGURE5}a, \ref{FIGURE5}b, \ref{FIGURE5}c) and \ref{FIGURE5}d), the trajectories converge to the respective stable fixed point  $\theta_c$, $\theta_c$, $\theta_d$ and $\theta_o$.} Finally, as shown in Figure \ref{FIGURE5}d, for initial conditions that are inside the magenta region, the pendulum will be attracted by the equilibrium point $\theta_d$.

One can notice that the basins of attractions of the system are dominant for initial values of $\theta$ and $\dot{\theta}$ in the vicinity of the equilibrium points. For large initial values of $\theta$ and $\dot{\theta}$, one can see a competition between the basin of attractions.

\section{Dynamical states when the coil magnets are powered by AC currents.}

$\quad$ We now analyze the situation where the coils are fed by sinusoidal current sources defined as $i_1  = I_{m1} \sin \left( {\omega t} \right)\;{\text{and}}\;i_2  = I_{m2} \sin \left( {\omega t - \varphi } \right)$. Angular frequency $\omega=2\pi f$, where $f$ is the frequency, and $\varphi$ represents the phase shift between the current sources. In this work, we will consider the case where the current sources are in phase ($\varphi=0$ rad) and the case where they are out of phase ($\varphi=\pm\pi$ rad).

\subsection{Bifurcation analysis}

\begin{figure}[t]
\begin{center}
\includegraphics[width=6.5cm,height=5.0cm]{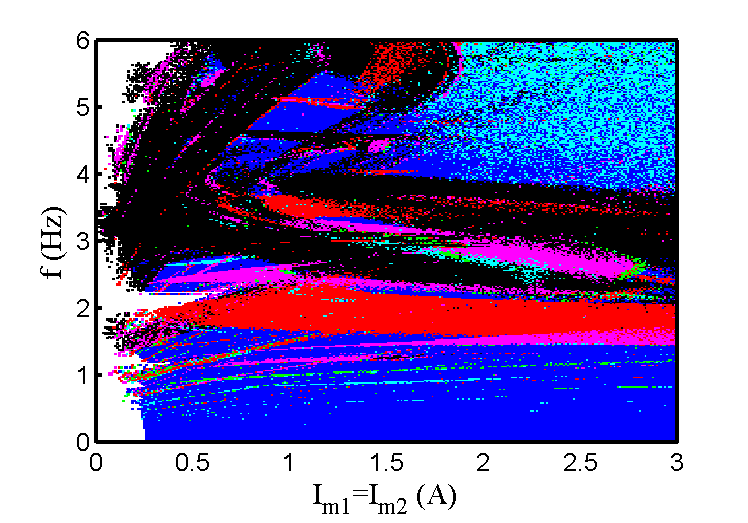}\hspace*{0.0cm}
\includegraphics[width=6.5cm,height=5.0cm]{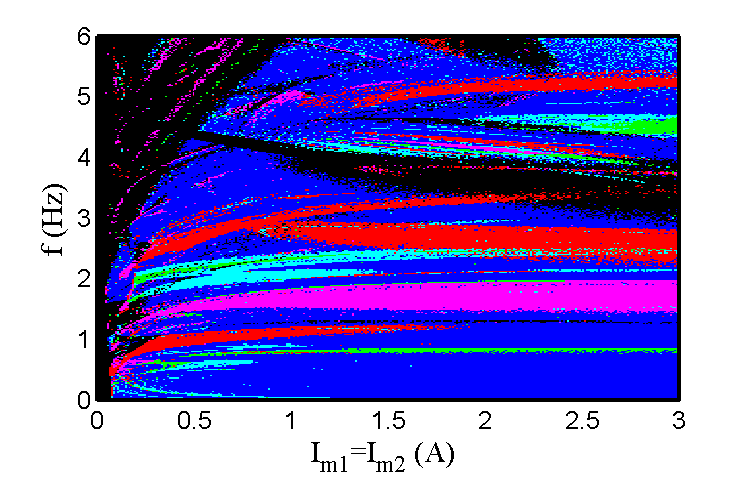}\\
\hspace*{0.5cm} a) \hspace*{6.5cm} b) \hfill\\
\caption{\label{FIGURE6}\footnotesize {Two-parameter bifurcation diagrams obtained when the current sources are a) in phase and b) out of phase in the plane $\left( {I_{m1}=I_{m2},\;f} \right)$ and for $\alpha=\frac{\pi}{10}$ rad.}}
\end{center}
\end{figure}

\begin{figure}[t]
\begin{center}
\includegraphics[width=6.5cm,height=5.0cm]{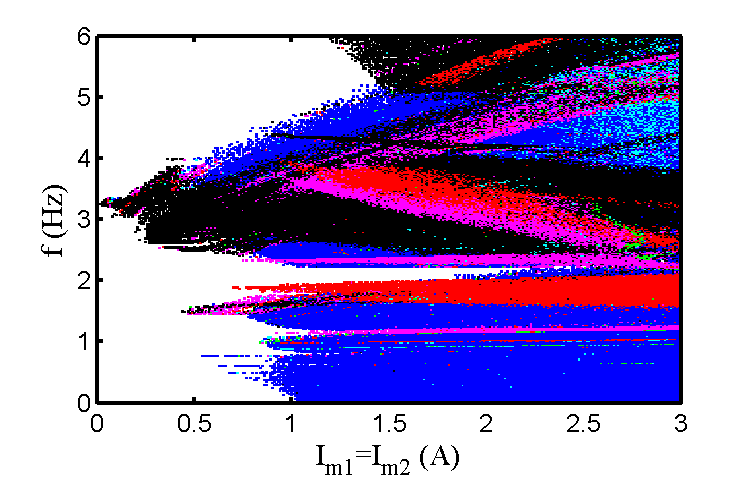}\hspace*{0.0cm}
\includegraphics[width=6.5cm,height=5.0cm]{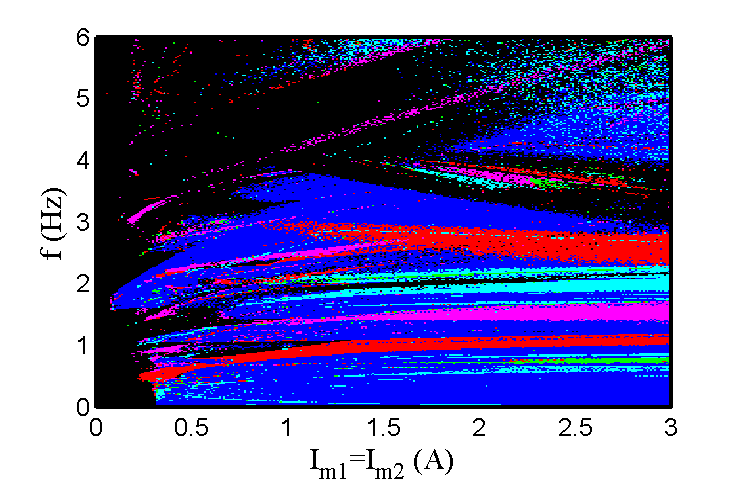}\\
\hspace*{0.5cm} a) \hspace*{6.5cm} b) \hfill\\
\caption{\label{FIGURE7}\footnotesize {Two-parameter bifurcation diagrams obtained when the current sources are a) in phase and b) out of phase in the plane $\left( {I_{m1}=I_{m2},\;f} \right)$ and for $\alpha=\frac{\pi}{6}$ rad.}}
\end{center}
\end{figure}

$\quad$ The richness of different dynamical behaviors exhibited by the system constituted of the magnetic pendulum and coils can be followed by the bifurcation diagrams. The investigated systems governed by the differential equation (\ref{EQUAT6}) have three control parameters $I_{m1}=I_{m2}$, $f$ and $\alpha$, and hence they are accessible experimentally, Figure \ref{FIGURE6} shows the two-parameter bifurcation diagrams highlighting the two control parameters $f$ and $I_{m1}=I_{m2}$  ($\alpha=\frac{\pi}{10}$ rad).
Similarly, Figure \ref{FIGURE7} shows the two-parameter bifurcation diagrams where the dependence of the system on two control parameters $f$ and $I_{m1}=I_{m2}$ ($\alpha=\frac{\pi}{6}$ rad) are employed.

Figures \ref{FIGURE6}a and \ref{FIGURE7}a are obtained when the current sources $i_1$ and $i_2$ are in phase ($\varphi=0$ rad). On the other hand, Figures \ref{FIGURE6}b and \ref{FIGURE7}b are constructed when the current sources $i_1$ and $i_2$ are out of phase ($\varphi=\pi$ rad).

To identify the dynamical behavior of the system, the common way is to determine the number $n$ in the period$-n$ response when varying the control parameters. To each period$-n$ response, with $n$, we have assigned a specific color. The white area denotes a region of no oscillation and the pendulum converges to one stable equilibrium point despite the time variation of the currents. Secondly, the black area, the magenta area, the red area, and the green area are attributed to the period-1, period-2, period-3, and period-4  trajectories, respectively. Thirdly, the cyan areas indicate a combination of period-5, period-6, period-7, period-8, period-9, and period-10. Finally, all responses with $n$ greater than $10$ and chaotic dynamics are marked by the blue area.

The bifurcation diagrams imply that the system does not perform oscillations for small amplitudes of the currents $i_1$ and $i_2$ being in phase. The region of no oscillation increases as the parameter $\alpha$ increases. When the currents $i_1$ and $i_2$ are out of phase, the pendulum performs period-1 oscillations for low current amplitudes. It should be emphasized that for frequencies above $4$ Hz, the system becomes very sensitive to the variations of its parameters, in particular the frequency and the amplitudes of the currents delivered by the coils.

\begin{figure}[h!]
\begin{center}
\includegraphics[width=6.5cm,height=4.0cm,angle=0]{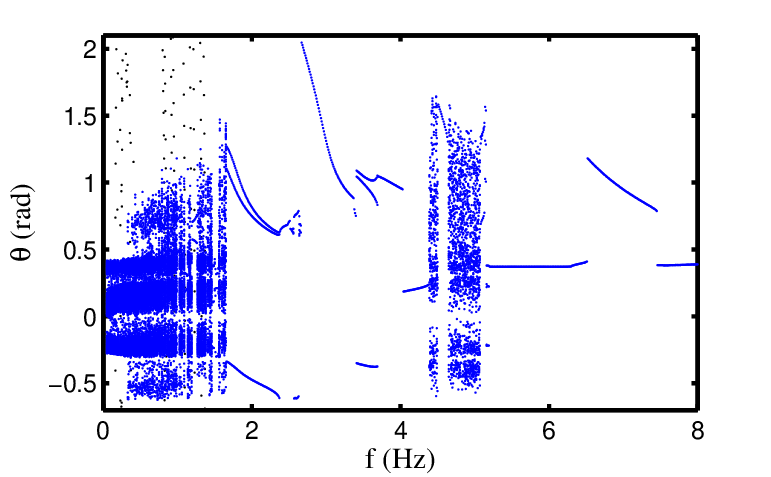}\hspace*{0.0cm}
\includegraphics[width=6.5cm,height=4.0cm,angle=0]{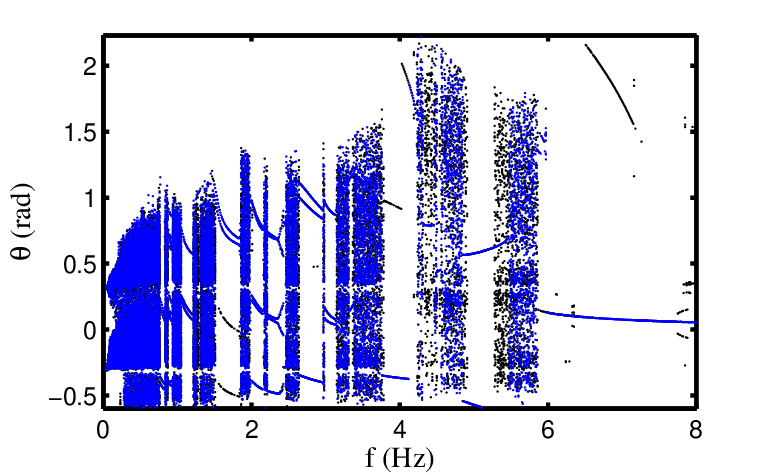}\\
\hspace*{0.70cm} a) \hspace*{6.0cm} b) \hfill\\
\includegraphics[width=6.5cm,height=3.0cm,angle=0]{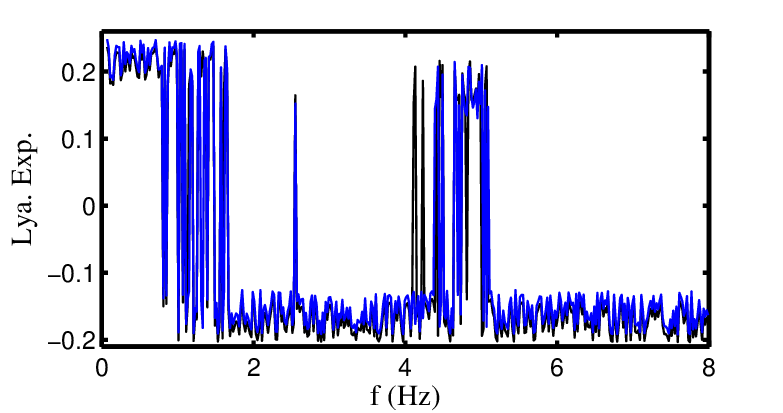}\hspace*{0.0cm}
\includegraphics[width=6.5cm,height=3.0cm,angle=0]{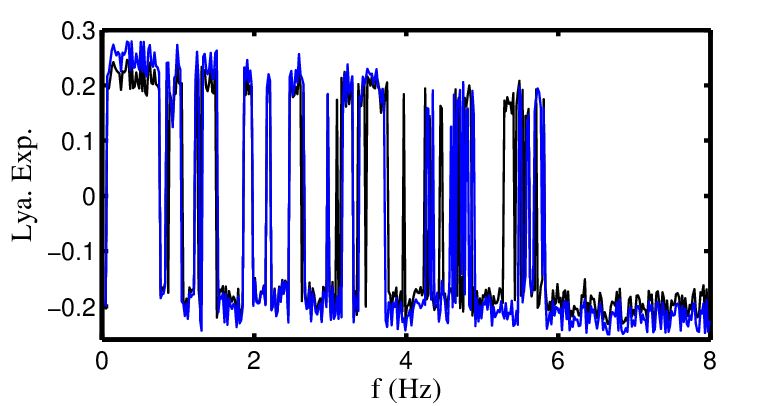}\\
\hspace*{0.70cm} c) \hspace*{6.0cm} d) \hfill\\
\caption{\label{FIGURE8}\footnotesize
Bifurcation diagrams and corresponding Lyapunov exponents as a function of the frequency $f$ obtained for a) and c) in phase and b) and d) out-of-phase coil currents respectively. The black curve is obtained with the following initial condition: $\theta(0)=0.57$ rad, $\dot \theta \left( 0 \right)=5.0$ rad/s. The blue curve is obtained for increasing $f$, where the final state at each iteration is used as the initial state for the next iteration.
}
\end{center}
\end{figure}

Furthermore, the effect of the frequency $f$ on the angular displacement $\theta$ of the pendulum is shown in Figures \ref{FIGURE8}a and \ref{FIGURE8}b for $\varphi=0$ rad and $\varphi=\pi$ rad, respectively, and fixed parameters $I_{m1}=I_{m2} =1$ A and $\alpha=\frac{\pi}{10}$.

\textcolor[rgb]{1.00,0.00,0.00}{The bifurcation diagrams displayed in Figure \ref{FIGURE8}a) and Figure \ref{FIGURE8}b) are obtained by plotting local maxima of the angular displacement $\theta$ as a function of the frequency $f$ of the external current sources, which increases in very small steps in the range $0$ Hz $\leq f \leq 7$ Hz. The corresponding Lyapunov exponents of Figure \ref{FIGURE8}a) and Figure \ref{FIGURE8}b) are plotted in Figure \ref{FIGURE8}c) and Figure \ref{FIGURE8}d) respectively. For the black curve, the initial conditions at each iteration of the control parameter $f$~are constant at $\theta(0)=-0.57$ rad, $\dot \theta \left( 0 \right)=5.0$ rad/s. Meanwhile, for the blue curve, the final state at each iteration is used as the initial state for the next iteration. As the graphs reveal, good agreements are found between the bifurcation diagrams and the corresponding graphs of the Lyapunov exponents.}

Based on the bifurcation diagrams presented in Figure \ref{FIGURE8}, it might be noticed that the angular displacement $\theta$ of the pendulum displays different dynamical behaviors as the periodicities and the amplitudes of the variable $\theta$ are not the same. Hence, solutions to equation (\ref{EQUAT6}) exhibit multistability features in different intervals of the control parameter.

\begin{figure}[h!]
\begin{center}
\includegraphics[width=6.5cm,height=5.0cm]{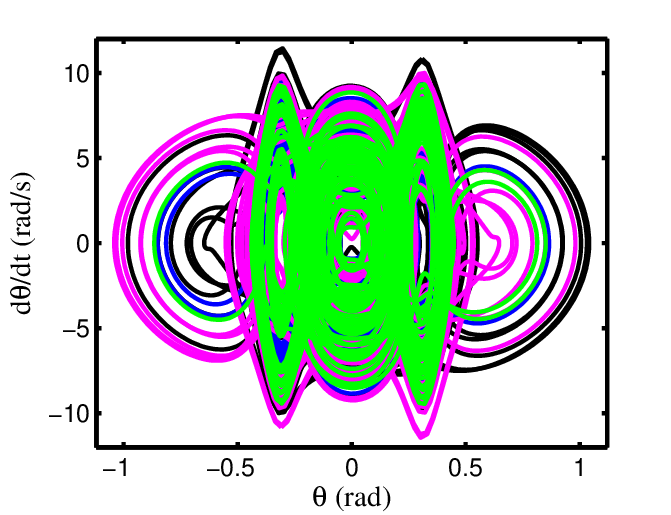}\hspace*{0.0cm}
\includegraphics[width=6.5cm,height=5.0cm]{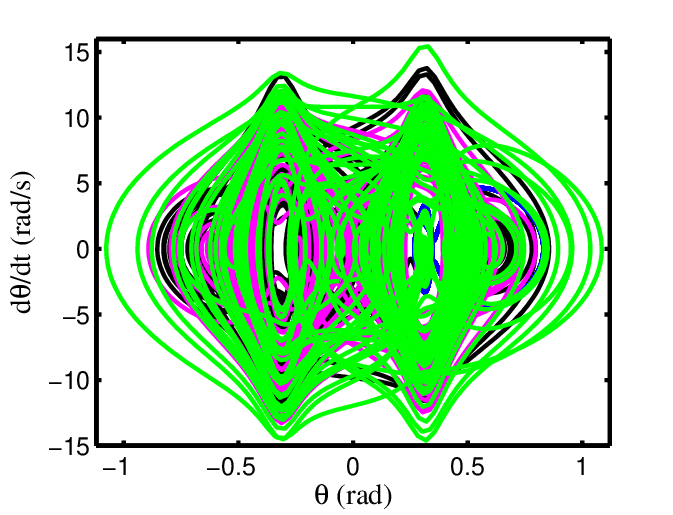}\\
\hspace*{0.50cm} a) \hspace*{6.0cm} b) \hfill\\
\caption{\label{FIGURE9}\footnotesize Phase portraits illustrating the multistability of the system for parameters: $f=1.35$ Hz, $\alpha=\frac{\pi}{10}$ rad and $I_{m1}=I_{m2}=1.0$ A. Subfigure a) is obtained for in-phase ($\varphi=0$ rad) coil currents, while b) corresponds to out of phase ($\varphi=\pi$ rad) coil currents.  For both figures, the trajectories are computed for different initial conditions.  The black curves ($\theta(0)=0.57$ rad,\; $\dot \theta \left( 0 \right)=8.14$ rad/s), blue curves ($\theta(0)=0.57$ rad,\; $\dot \theta \left( 0 \right)=-8.14$ rad/s), magenta curves ($\theta(0)=-0.57$ rad,\; $\dot \theta \left( 0 \right)=-8.14$ rad/s), and green curves ($\theta(0)=-0.57$ rad,\; $\dot \theta \left( 0 \right)=8.14$ rad/s).}
\end{center}
\end{figure}

A confirmation of the multistable behavior of the system is presented in Figure \ref{FIGURE9} where one sees that for the same parameters different trajectories can be obtained depending on the values of the initial solutions.  Figure \ref{FIGURE9}a) presents a coexistence of four chaotic behaviors when the currents are in phase and $f=1.35$ Hz. A similar behavior is presented in Figure \ref{FIGURE9}b) plotted for $f=1.35$ Hz and with the current sources out of phase, one sees a coexistence of four chaotic behaviors. The curves in black are obtained by using the initial conditions $\theta(0)=0.57$ rad and $\dot \theta \left( 0 \right)=8.14$ rad/s, the curves in blue are obtained by using the initial conditions $\theta(0)=0.57$ rad, $\dot \theta \left( 0 \right)=-8.14$ rad/s, whereas the curves in magenta are obtained for the initial conditions $\theta(0)=-0.57$ rad, $\dot \theta \left( 0 \right)=-8.14$ rad/s. Finally, the curves in green are obtained with the following initial conditions $\theta(0)=-0.57$ rad and $\dot \theta \left( 0 \right)=8.14$ rad/s.

\subsection{Numerical and analytical analysis}

\begin{figure}[h!]
\begin{center}
\includegraphics[width=5.0cm,height=4.3cm]{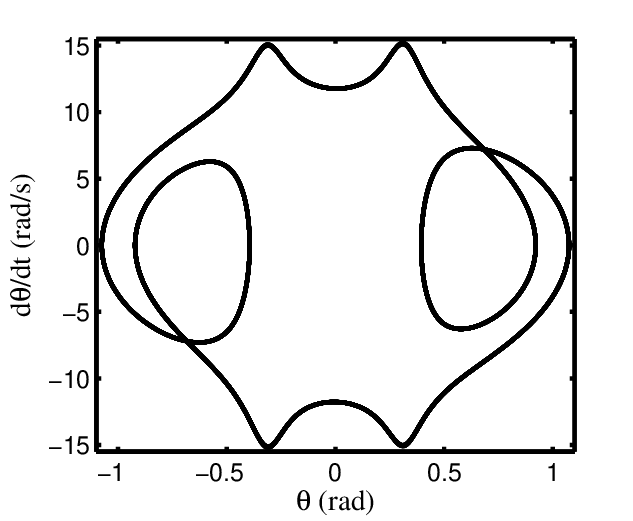}\hspace*{0.5cm}
\includegraphics[width=7.0cm,height=4.3cm]{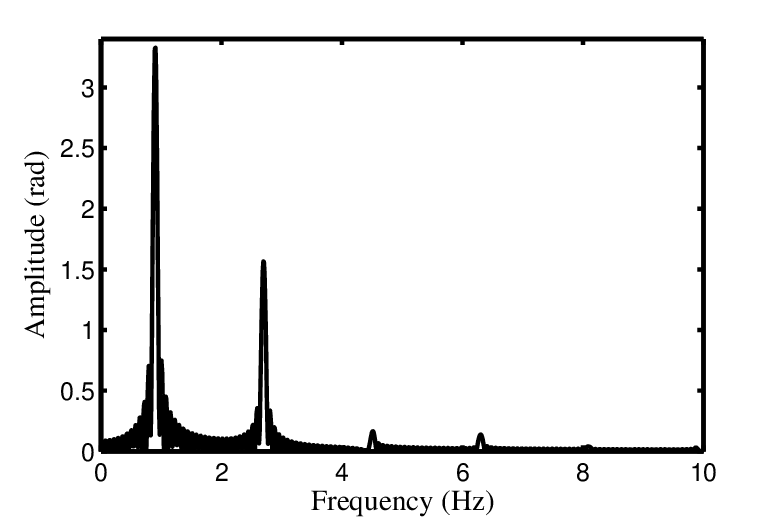}\\
\hspace*{0.0cm} a) \hspace*{6.5cm} b) \hfill\\
\caption{\label{FIGURE10}\footnotesize {a) Phase portrait of the system when $f=1.8$ Hz, $I_{m1}=I_{m2}=1$ A, $\varphi=0$ rad and $\alpha=\frac{\pi}{10}$ rad. b) Corresponding frequency spectrum of angular displacement.}}
\end{center}
\end{figure}

Bifurcation diagrams shown in Figures \ref{FIGURE6}a) and \ref{FIGURE7}a), present domains/ranges of parameters where the system exhibits periodic solutions. For instance, in Figure \ref{FIGURE6}a), period-1 oscillations exist for frequencies close to $3$ Hz and for some range of current. In the same Figures \ref{FIGURE6}a) and \ref{FIGURE7}a), period-3 oscillations are present around the frequency $1.8$ Hz or $2$ Hz for current amplitude above $0.5$ A (below this value, there is an alternation of different dynamical states). For Figures \ref{FIGURE6}b) and \ref{FIGURE7}b), period-1 oscillations for small values of the current amplitude (close to $0$ mA) and for any frequency are observed. Therefore, numerical methods might be employed to determine the angular amplitude of oscillations and plot its amplitude versus the current amplitude or the frequency depending on the case considered. We have done this in the approximate form of the pendulum differential equation. We start with a situation where period-3 oscillations exist. Figure \ref{FIGURE10}a) shows the phase portrait of the system for the following fixed parameters: $f=1.8$ Hz, $I_{m1}=I_{m2}=1$ A, $\varphi=0$ rad and $\alpha=\frac{\pi}{10}$. The corresponding Fourier transform amplitude spectrum is presented in Figure \ref{FIGURE10}b).

\begin{figure}[h!]
\begin{center}
\includegraphics[width=5.0cm,height=4.3cm]{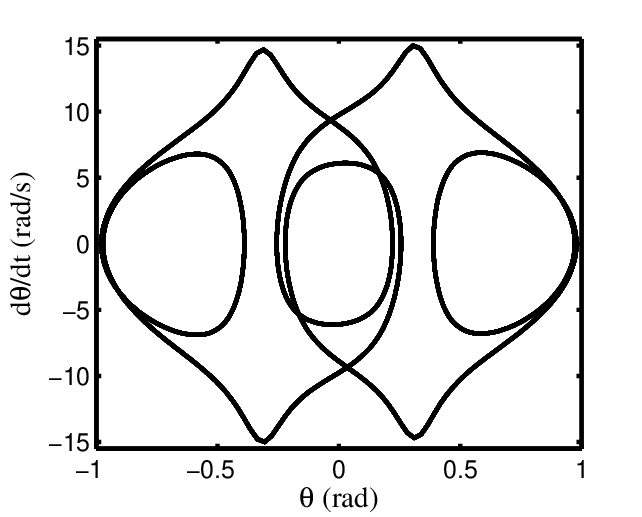}\hspace*{0.5cm}
\includegraphics[width=7.0cm,height=4.3cm]{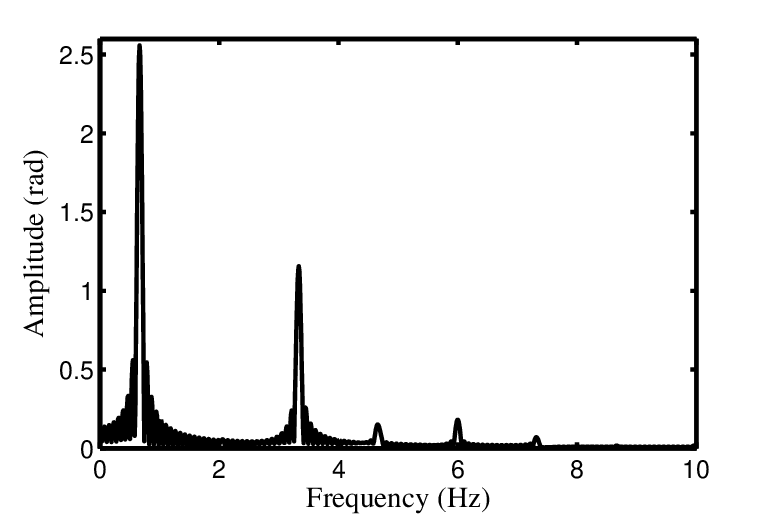}\\
\hspace*{0.0cm} a) \hspace*{6.0cm} b) \hfill\\
\caption{\label{FIGURE11}\footnotesize {a) Phase portrait of the system when $f=2.0$ Hz, $I_{m1}=I_{m2}=1$ A, $\varphi=\pi$ rad and $\alpha=\frac{\pi}{10}$ rad. b) Corresponding frequency spectrum of angular displacement.}}
\end{center}
\end{figure}

\begin{figure}[h!]
\begin{center}
\includegraphics[width=5.0cm,height=4.3cm]{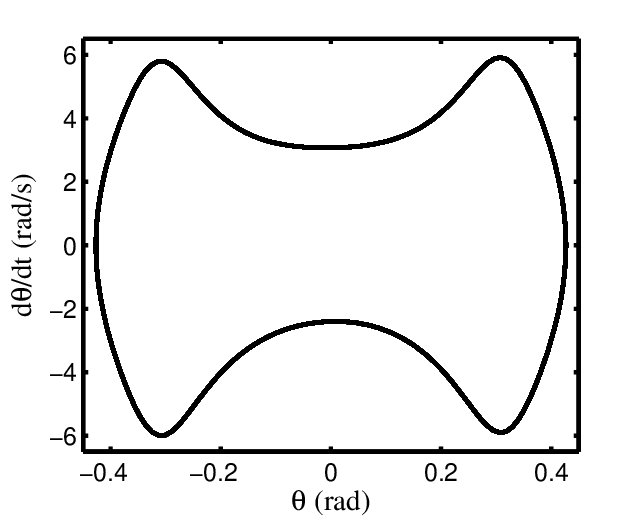}\hspace*{0.5cm}
\includegraphics[width=7.0cm,height=4.3cm]{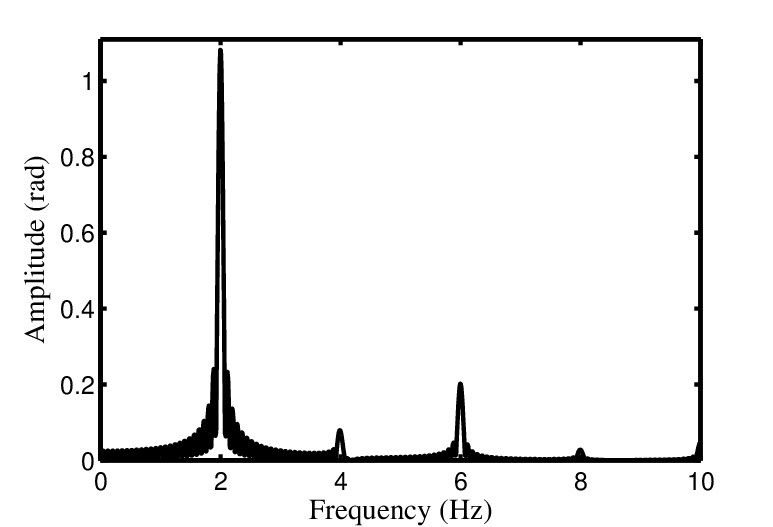}\\
\hspace*{0.0cm} a) \hspace*{6.0cm} b) \hfill\\
\caption{\label{FIGURE12}\footnotesize {a) Phase portrait of the system when $f=2.0$ Hz, $I_{m1}=I_{m2}=400$ mA, $\varphi=\pi$ rad and $\alpha=\frac{\pi}{10}$ rad. b) Corresponding frequency spectrum of angular displacement.}}
\end{center}
\end{figure}

Namely, Figure \ref{FIGURE10}b) shows that dominant frequency components are $0.9$ Hz and $2.7$ Hz, which correspond to $\frac{1}{2}f$ and $\frac{3}{2}f$, respectively. As shown in Figure \ref{FIGURE11}, the situation is different if the currents $i_1$ and $i_2$ are out of phase.
Figure \ref{FIGURE11} is obtained for $f=2.0$ Hz, $I_{m1}=I_{m2}=1$ A, $\varphi=\pi$ rad and $\alpha=\frac{\pi}{8}$ rad. As indicated in the corresponding frequency spectrum, the important frequency components are $0.66$ Hz, $2.0$ Hz, and $3.33$ Hz. These frequencies are related to the frequency of the external sources, that is, $\frac{1}{3}f$, $\frac{3}{3}f$, and $\frac{5}{3}f$, respectively. For some sets of the system parameters, we found that the important frequency component is $f$, as illustrated in the graphs in Figure \ref{FIGURE12}.

Figure \ref{FIGURE12} is plotted for $f=2.0$ Hz, $I_{m1}=I_{m2}=400$ mA, $\varphi=\pi$ rad and $\alpha=\frac{\pi}{10}$ rad. \textcolor[rgb]{1.00,0.00,0.00}{As indicated in the corresponding frequency spectrum, the crucial roles in the system dynamics are played by the frequency components $f$ and $3f$}.

In order to carry out analytical investigations, the differential equation (\ref{EQUAT6}) can be recast to the following form \begin{equation}\label{EQUAT13ABC}
J\frac{{d^2 \theta }}{{dt^2 }} + \left( {\beta  + \chi M_s } \right)\frac{{d\theta }}{{dt}} + C\theta  + \gamma \left( {\theta  - \frac{1}{6}\theta ^3 } \right) = \frac{{\eta \left( {\theta  - \alpha } \right)i_1 }}
{{\left( {\theta  - \alpha } \right)^2  + \mu ^2 }} + \frac{{\eta \left( {\theta  + \alpha } \right)i_2 }}
{{\left( {\theta  + \alpha } \right)^2  + \mu ^2 }},
\end{equation}
where $\Gamma_s$ and $\sin \theta$ have been replaced by their corresponding first and third-order series expansions, respectively. The external torque has been approximated by ratios of polynomial functions, and the coefficient $\mu$ and $\eta$ are defined as
\begin{equation}\label{EQUAT14ABC}
\begin{gathered}
  \mu  = \pi  - \arccos \left( {1 + \varepsilon  - \sqrt {\varepsilon ^2  + 2\varepsilon  + 4} } \right), \hfill \\
  \eta  = \frac{{9\mu \sigma \sqrt {1 - \left( {1 + \varepsilon  - \sqrt {\varepsilon ^2  + 2\varepsilon  + 4} } \right)^2 } }}{{\left( {2\varepsilon  + 2 - \sqrt {\varepsilon ^2  + 2\varepsilon  + 4} } \right)^{\frac{3}
{2}} }}. \hfill \\\end{gathered}
\end{equation}
Owing to the value of the parameter $\varphi$, there are two different cases. In this section, we consider the situation where the currents $i_1$ and $i_2$ are in phase ($\varphi=0$ rad). Taking advantage of the results obtained previously, and using the balanced harmonic method, we consider the solution of equation (\ref{EQUAT13ABC}) in the form
\begin{equation}\label{EQUAT15}
\theta  = A\cos \left( {\frac{{\omega t}}{2}} \right) + B\sin \left( {\frac{{\omega t}}{2}} \right).
\end{equation}
At this level, the quantity $\theta _m  = \sqrt {A^2  + B^2 }$ is the amplitude of the analytical signal and represents our unknown variable. The substitution of equation (\ref{EQUAT15}) into the differential equation (\ref{EQUAT13ABC}), and after some transformations, the amplitude $\theta_m$ is solution of the following twelfth order polynomial
\begin{equation}\label{EQUAT16}
A_{12} \theta _m^{12}  + A_{10} \theta _m^{10}  + A_8 \theta _m^8  + A_6 \theta _m^6  + A_4 \theta _m^4  + A_2 \theta _m^2  + A_0  = 0,
\end{equation}
where
\begin{equation}\label{EQUAT17}
\begin{gathered}
  A_{12}  = 1225\gamma ^2 ,\;\;A_{10}  = 5600\gamma \left( {\upsilon \gamma  + 3\psi } \right), \hfill \\
  A_8  = 32\left[ {72\beta '^2  + 5\gamma \left( {61\upsilon ^2  + 84\alpha ^2 \upsilon  + 84\alpha ^4 } \right) + 60\psi \left( {41\upsilon \gamma  + 30\psi } \right)} \right], \hfill \\  A_6  = 256\left[ {72\upsilon \left( {\beta '^2  + 15\psi ^2 } \right) + 30\gamma \upsilon \left( {\upsilon  + 2\alpha ^2 } \right)^2  + 15\gamma ^2 \psi \left( {37\upsilon ^2  + 52\upsilon \alpha ^2  + 52\alpha ^4 } \right)} \right], \hfill \\  A_4  = 73728\left[ {\beta '^2 \left( {\upsilon ^2  + 2\upsilon \alpha ^2  + 2\alpha ^4 } \right) + \psi ^2 \left( {7\upsilon ^2  + 10\upsilon \alpha ^2  + 10\alpha ^4 } \right)} \right] +  \hfill \\  \quad \quad 256\left( {\upsilon  + 2\alpha ^2 } \right)^2 \left[ {9\gamma ^2 \left( {\upsilon  + 2\alpha ^2 } \right)^2  + 456\gamma \psi \upsilon } \right], \hfill \\  A_2  = 36864\left( {\upsilon  + 2\alpha ^2 } \right)^2 \left[ {4\upsilon \left( {\beta '^2  + 3\psi ^2 } \right) + \gamma \psi \left( {\upsilon  + 2\alpha ^2 } \right)^2 } \right], \hfill \\  A_0  =  - \upsilon ^2 \eta ^2 \left[ {\left( {I_{m1}  + I_{m2} \cos \varphi } \right)^2  + I_{m2}^2 \sin ^2 \varphi } \right] + 4\left( {\beta '^2  + \psi ^2 } \right)\left( {\upsilon  + 2\alpha ^2 } \right)^4 . \hfill \\\end{gathered}
\end{equation}
The newly introduced parameters $\beta ' = \left( {\beta  + \chi M_s } \right)\omega$, $\upsilon ={\mu ^2  - \alpha ^2 }$ and $\psi  = J\omega ^2  - C - \gamma$ are used to simplify the notations. In the latter case, the opposite current sources do not affect the amplitude of the pendulum.

\begin{figure}[t]
\begin{center}
\includegraphics[width=6.5cm,height=4.5cm]{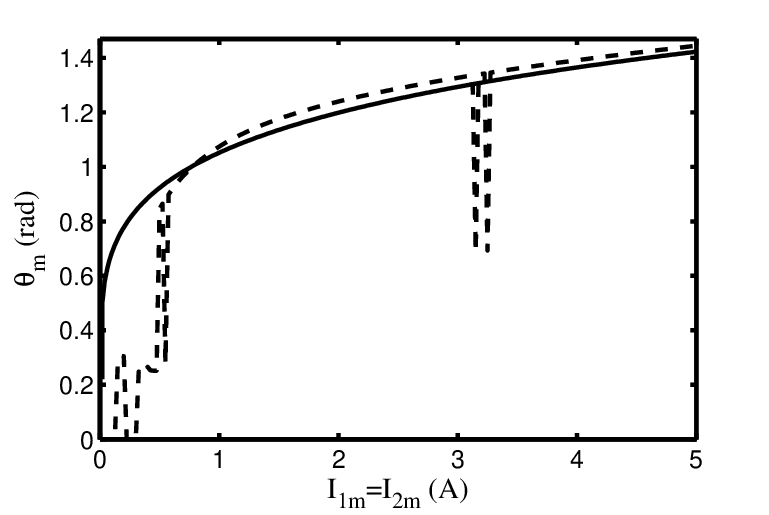}\hspace*{0.0cm}
\includegraphics[width=6.5cm,height=4.5cm]{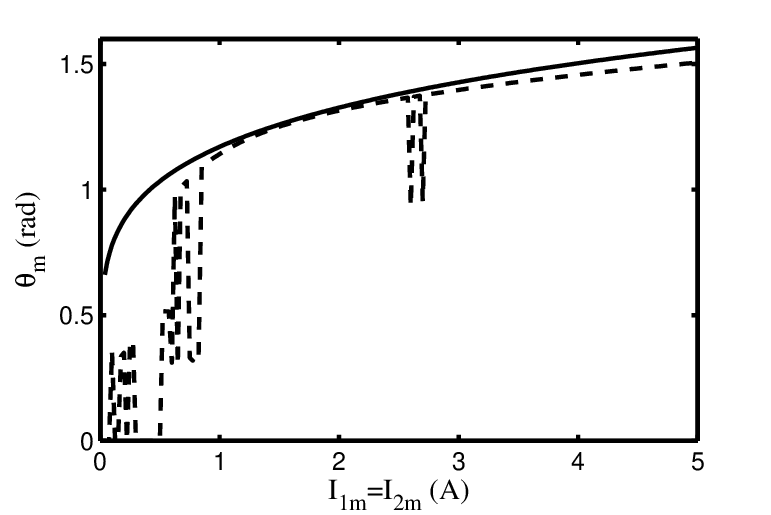}\\
\hspace*{1.5cm} a) \hspace*{6.0cm} b) \hfill\\
\caption{\label{FIGURE13}\footnotesize {Amplitude of angle $\theta_m$ versus the magnitudes $I_{m1}= I_{m2}$ of the current sources for $f=1.8$ Hz and $\varphi=0$ rad. a) $\alpha=\frac{\pi}{10}$ rad and b) $\alpha=\frac{\pi}{8}$ rad. The curves with solid lines correspond to analytical results, while the curves with dashed lines are obtained numerically.}}
\end{center}
\end{figure}

To verify our analysis, we use the magnitudes $I_{m1} = I_{m2}$ of the current sources as the control parameter, and the system is analyzed for two different angles $\alpha$. Figure \ref{FIGURE13}a) and Figure \ref{FIGURE13}b) are plotted for $\alpha=\frac{\pi}{10}$ rad and $\alpha=\frac{\pi}{8}$ rad, respectively. The frequency used in Figure \ref{FIGURE13} is $f=1.8$ Hz and the currents are in phase.

\begin{figure}[t]
\begin{center}
\includegraphics[width=5.5cm,height=4.5cm]{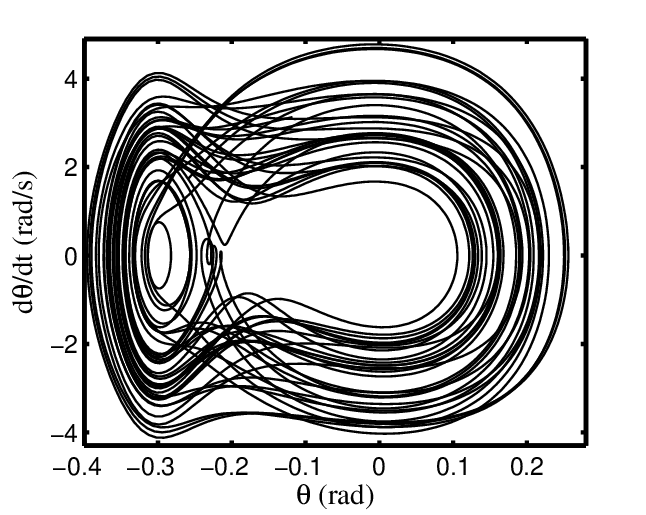}\hspace*{0.0cm}
\includegraphics[width=7.0cm,height=4.5cm]{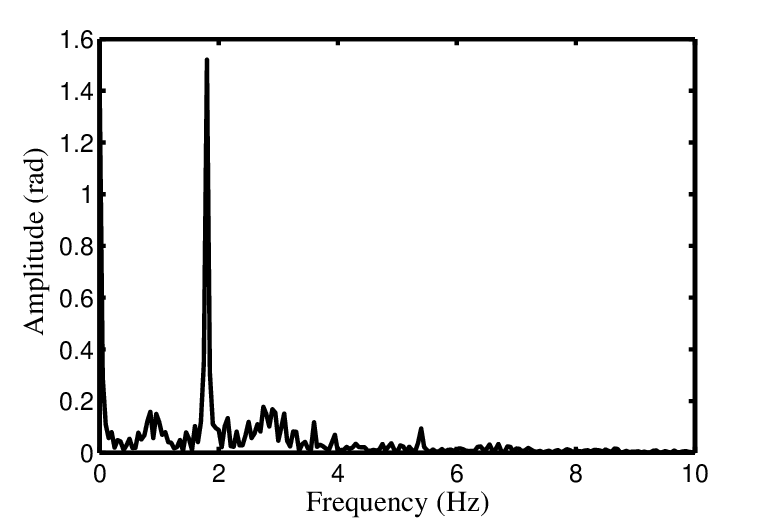}\\
\hspace*{0.5cm} a) \hspace*{7.0cm} b) \hfill\\
\caption{\label{FIGURE14}\footnotesize {Bifurcation diagrams as a function of the amplitude $I_{m1}=I_{m2}$ and plotted for different values of $\alpha$: a) $\alpha=\frac{\pi}{10}$ rad and b) $\alpha=\frac{\pi}{8}$ rad.}}
\end{center}
\end{figure}

\begin{figure}[t]
\begin{center}
\includegraphics[width=5.5cm,height=4.5cm]{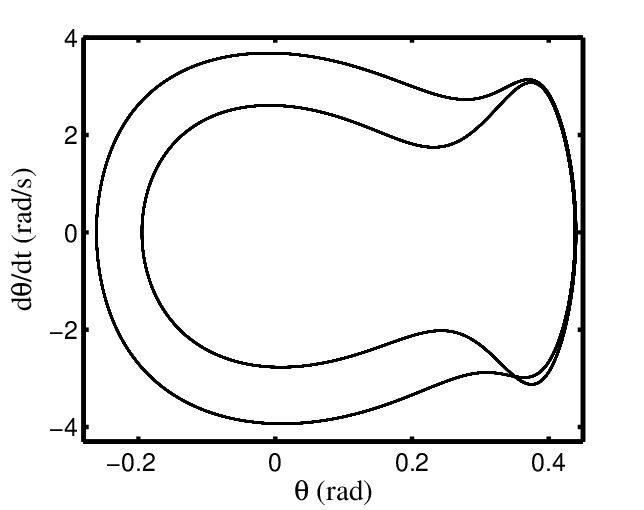}\hspace*{0.0cm}
\includegraphics[width=7.0cm,height=4.5cm]{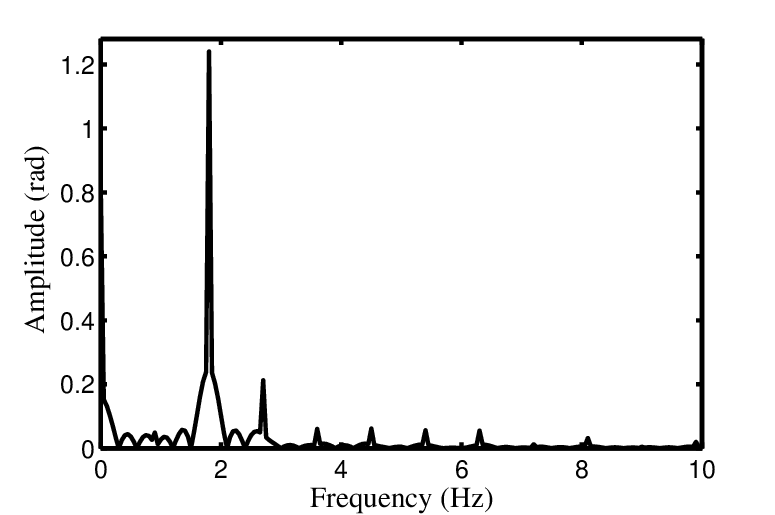}\\
\hspace*{0.5cm} a) \hspace*{7.0cm} b) \hfill\\
\caption{\label{FIGURE15}\footnotesize {Bifurcation diagrams as a function of the amplitude $I_{m1}=I_{m2}$ and plotted for different values of $\alpha$: a) $\alpha=\frac{\pi}{10}$ rad and b) $\alpha=\frac{\pi}{8}$ rad.}}
\end{center}
\end{figure}

The analytical/numerical solutions are plotted with solid/dashed lines. Figures \ref{FIGURE13}a) and \ref{FIGURE13}b) indicate excellent agreement between our analytical and simulation results when the current amplitudes are greater than $0.5$ A and $0.8$ A, respectively. \textcolor[rgb]{1.00,0.00,0.00}{This can be understood since as presented in Figure \ref{FIGURE6}, the responses of the system are dominated by {period-3} oscillations. Equation (\ref{EQUAT15}) implies that $\frac{1}{2}f$ is the important frequency component of the response signal. In the other regions of Figures \ref{FIGURE13}a) and \ref{FIGURE13}b) ($I_m<0.5$ A and $I_m<0.8$ A), the alternation of different dynamical states including chaotic oscillations is observed. This is illustrated in the phase portraits of Figure \ref{FIGURE14}a) and Figure \ref{FIGURE15}a) plotted for $\varphi=\frac{\pi}{10}$ rad and $\varphi=\frac{\pi}{8}$ rad respectively.}

\textcolor[rgb]{1.00,0.00,0.00}{Figures \ref{FIGURE14}a) and \ref{FIGURE15}a) are obtained for $f=1.8$ Hz and $I_{m1}=I_{m2}=200$ mA. As indicated in the corresponding frequency spectrum presented in Figures \ref{FIGURE14}b) and \ref{FIGURE15}b), the important frequency components are $0.0$ Hz, $1.8$ Hz, $2.7$ Hz, and $3.6$ Hz. These frequencies are related to the frequency of the external sources as $\frac{0}{2}f$, $\frac{2}{2}f$, $\frac{3}{2}f$, and $\frac{4}{2}f$, respectively. Consequently, the approximated analytical solution for $I_m<0.5$ A and $I_m<0.8$ A will have the following form
\begin{equation}\label{EQUAT15A}
\theta  =\theta_0+ A_1\cos \left( {{\omega t}} \right) + B_1\sin \left( {{\omega t}} \right).
\end{equation}}

It should be noticed that for certain small values of the current magnitudes, and hence the pendulum comes to rest after the transient process.

Similar analysis can be performed when the currents $i_1$ and $i_2$ are out of phase. But one has to notice that the important frequency component is generally $\frac{1}{2}f$ when the current sources are in phase. In contrast, when the current sources are out of phase, the significant frequency component is $\frac{1}{3}f$ and sometimes $\frac{3}{3}f$.

\section{Laboratory}

\subsection{Experimental rig}

\begin{figure}[h!]
\begin{center}
\includegraphics[width=4.5cm,height=5.5cm]{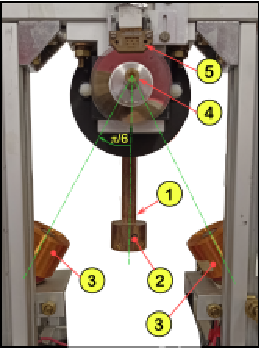}\hspace*{2.0cm}
\includegraphics[width=5.4cm,height=3.0cm]{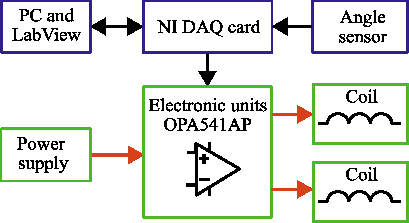}\\
\hspace*{0.5cm} a) \hspace*{7.0cm} b) \hfill\\
\caption{\label{FIGURE16}\footnotesize {Experimental setup of magnetic pendulum subjected to two electric coils (a), where: 1-pendulum, 2-neodymium magnet, 3-electric coil, 4-brass shaft, 5-incremental optical encoder.}}
\end{center}
\end{figure}

The experimental rig is presented in Figure \ref{FIGURE16}a) and consists of a pendulum {(1)} with a neodymium magnet {(2)} attached at the tip. Electric coils {(3)} are placed symmetrically on both sides of the resting pendulum at an angle of $30$ degrees.

The pendulum is suspended on a brass shaft {(4)} supported by ball bearings. The angular position measurement is performed using an incremental optical encoder {(5)}. All construction is made of non-magnetic materials like textile composites, aluminum, brass, and polymer materials. Thanks to that the magnet-coils interaction is protected from external undesired magnetic effects like induction of eddy currents or additional magnetic fields of some ferromagnetic parts.

The electric coils characterized by $22$ mH of inductance and $10.6$ $\mathrm{\Omega}$ of resistance are powered by two electronic systems based on OPA541AP operational amplifiers. The signal flow diagram in the system is shown in Figure \ref{FIGURE16}b). The task of the electronic systems is to convert the voltage signal coming from the NI DAQ card into a current signal flowing in the coils. The voltage signal is generated in the LabView program, which is also responsible for displaying the information collected by the HEDS-6140-B13 angle position sensor. The source of electricity powering the coils is the KORAD 3305D power supply. Due to the maximum current and voltage efficiency (30V 5A for two channels), we were not able to obtain experimentally all the numerical results contained in this paper.

\subsection{Experimental vs. numerical results}

By varying the frequency and amplitudes of the current sources while keeping the angle $\alpha$ constant at $\alpha=\frac{\pi}{6}$ rad, the simulation results using the two models described earlier are compared to the experimental ones.

\begin{figure}[b]
\begin{center}
\includegraphics[width=4.3cm,height=3.8cm]{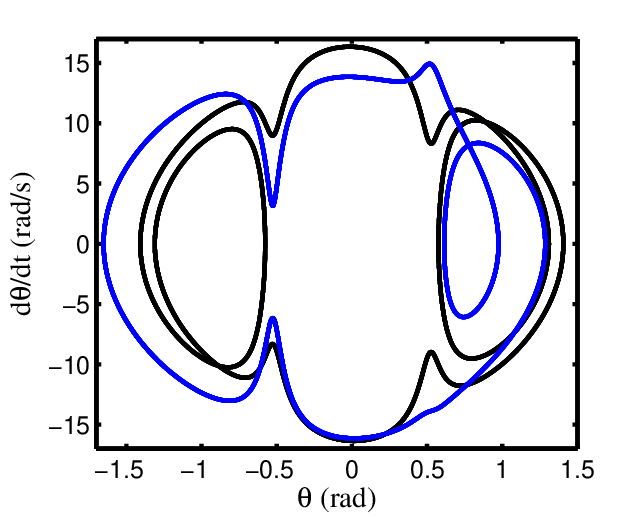}\hspace*{0.0cm}
\includegraphics[width=4.3cm,height=3.8cm]{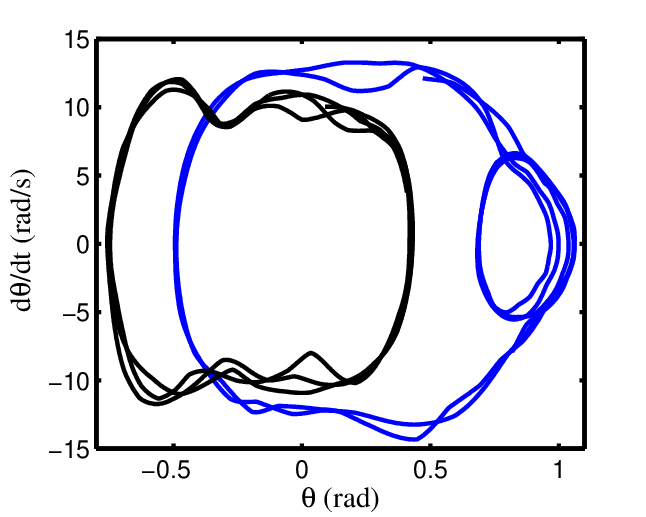}\hspace*{0.0cm}
\includegraphics[width=4.3cm,height=3.8cm]{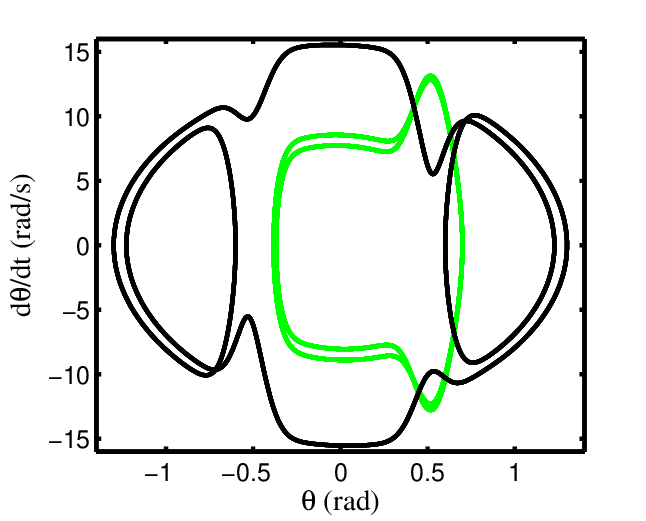}\\
\hspace*{1.0cm} a) \hspace*{4.0cm} b) \hspace*{4.0cm} c) \hfill\\
\caption{\label{FIGURE17}\footnotesize {Different phase portraits of the system. a) Obtained numerically using the first model (\ref{EQUAT6}), b) obtained experimentally, and c) obtained numerically using the second model (\ref{EQUAT6A}), for: $I_{m1}=I_{m2}=1.5$ A, $f=3.5$ Hz, $\alpha=\frac{\pi}{6}$ rad and $\varphi=0$ rad. Black curves are plotted for: $\theta(0)=0.57$ rad,\; $\dot \theta \left( 0 \right)=8.14$ rad/s. Blue and green curves are plotted for: $\theta(0)=0.0$ rad,\; $\dot \theta \left( 0 \right)=0.14$ rad/s and $\theta(0)=0.232$ rad,\; $\dot \theta \left( 0 \right)=-9.968$ rad/s, respectively.}}
\end{center}
\end{figure}

\begin{figure}[t]
\begin{center}
\includegraphics[width=4.3cm,height=3.8cm]{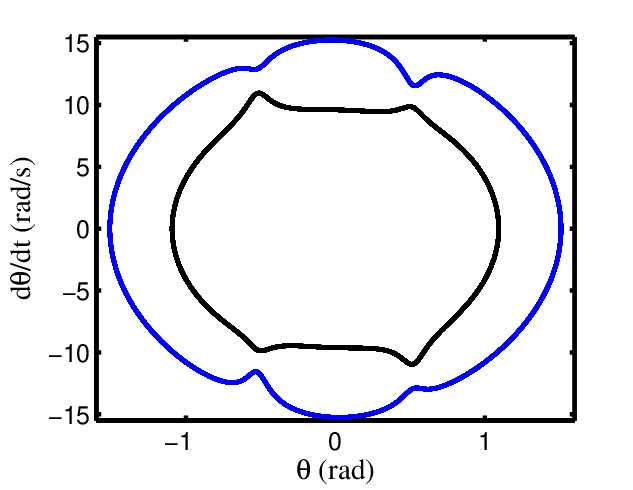}\hspace*{0.0cm}
\includegraphics[width=4.3cm,height=3.8cm]{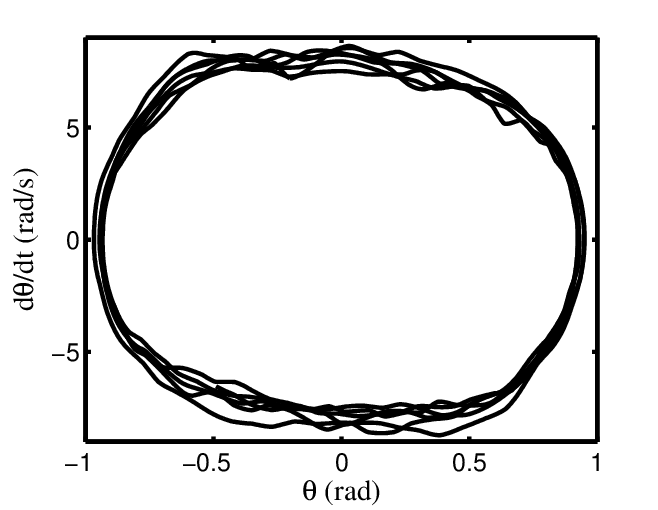}\hspace*{0.0cm}
\includegraphics[width=4.3cm,height=3.8cm]{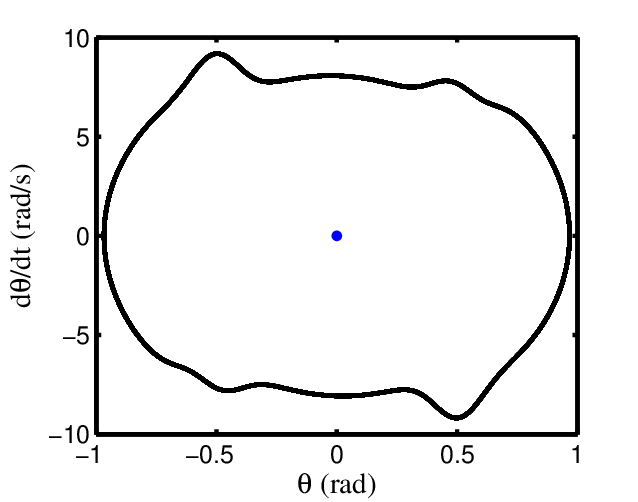}\\
\hspace*{1.0cm} a) \hspace*{4.0cm} b) \hspace*{4.0cm} c) \hfill\\
\caption{\label{FIGURE18}\footnotesize {Different phase portraits of the system. a) Obtained numerically using the first model (\ref{EQUAT6}), b) obtained experimentally, and c) obtained numerically using the second model (\ref{EQUAT6A}), for: $I_{m1}=I_{m2}=500$ mA, $f=3.0$ Hz, $\alpha=\frac{\pi}{6}$ rad and $\varphi=0$ rad. Black curves are plotted for: $\theta(0)=0.57$ rad,\; $\dot \theta \left( 0 \right)=8.14$ rad/s. Blue and green curves are plotted for: $\theta(0)=0.0$ rad,\; $\dot \theta \left( 0 \right)=0.14$ rad/s and $\theta(0)=0.232$ rad,\; $\dot \theta \left( 0 \right)=-9.968$ rad/s, respectively.}}
\end{center}
\end{figure}

\begin{figure}[t]
\begin{center}
\includegraphics[width=4.3cm,height=3.8cm]{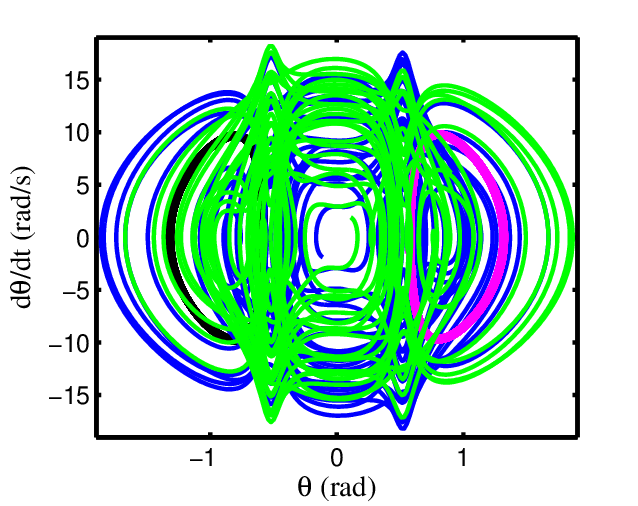}\hspace*{0.0cm}
\includegraphics[width=4.3cm,height=3.8cm]{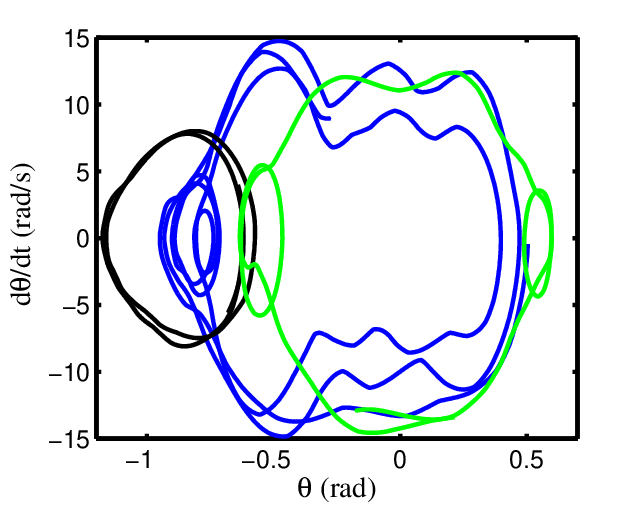}\hspace*{0.0cm}
\includegraphics[width=4.3cm,height=3.8cm]{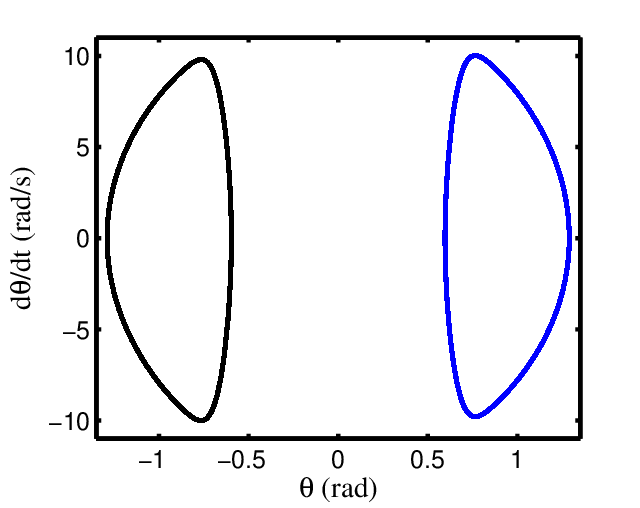}\\
\hspace*{1.0cm} a) \hspace*{4.0cm} b) \hspace*{4.0cm} c) \hfill\\
\caption{\label{FIGURE19}\footnotesize {Different phase portraits of the system. a) Obtained numerically using the first model (\ref{EQUAT6}), b) obtained experimentally, and c) obtained numerically using the second model (\ref{EQUAT6A}), for: $I_{m1}=I_{m2}=1.5$ A, $f=4.0$ Hz, $\alpha=\frac{\pi}{6}$ rad and $\varphi=0$ rad. Black and blue curves are plotted for: $\theta(0)=0.57$ rad,\; $\dot \theta \left( 0 \right)=8.14$ rad/s and $\theta(0)=0.57$ rad,\; $\dot \theta \left( 0 \right)=-8.14$ rad/s, respectively. Magenta and green curves  are plotted for: $\theta(0)=-0.57$ rad,\; $\dot \theta \left( 0 \right)=-8.14$ rad/s and $\theta(0)=-0.57$ rad,\; $\dot \theta \left( 0 \right)=8.14$ rad/s, respectively. 
}}
\end{center}
\end{figure}

We first consider the case where the current sources are in phase ($\varphi=0$ rad) and the results are presented in Figures \ref{FIGURE17}, \ref{FIGURE18} and \ref{FIGURE19}. Figures \ref{FIGURE17}a), \ref{FIGURE18}a) and \ref{FIGURE19}a) show the simulation results obtained from the first model, while Figures \ref{FIGURE17}c), \ref{FIGURE18}c) and \ref{FIGURE19}c) are attributed to simulation results of the second model. For comparison, the experimental results are given in Figures \ref{FIGURE17}b), \ref{FIGURE18}b) and \ref{FIGURE19}b).

\clearpage

As one can notice from the graphs, the multistable behavior observed theoretically is confirmed by the experimental results.

\begin{figure}[h!]
\begin{center}
\includegraphics[width=4.3cm,height=3.8cm]{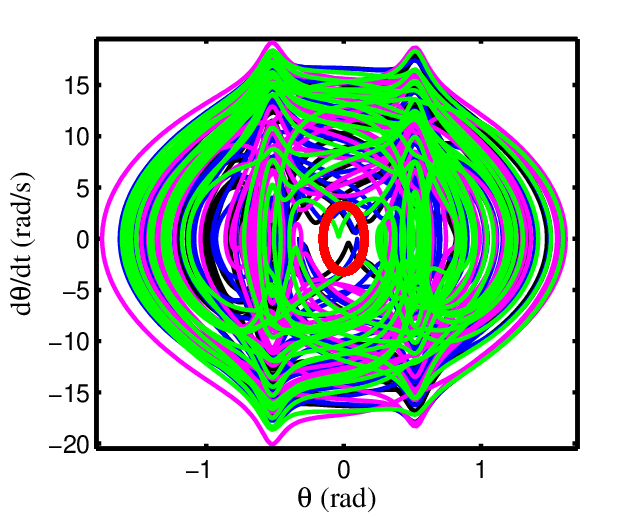}\hspace*{0.0cm}
\includegraphics[width=4.3cm,height=3.8cm]{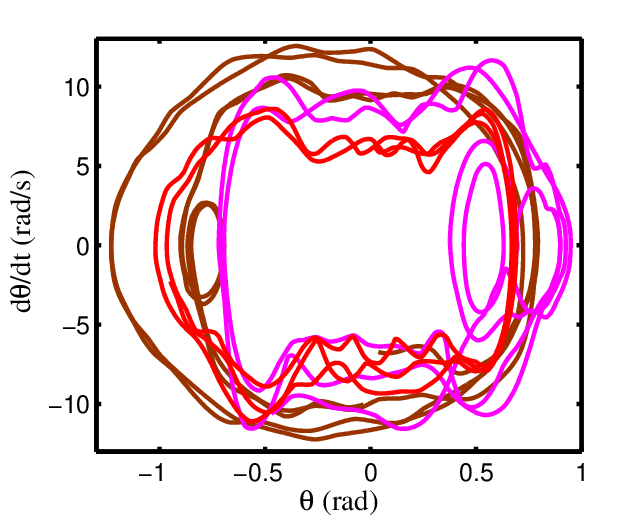}\hspace*{0.0cm}
\includegraphics[width=4.3cm,height=3.8cm]{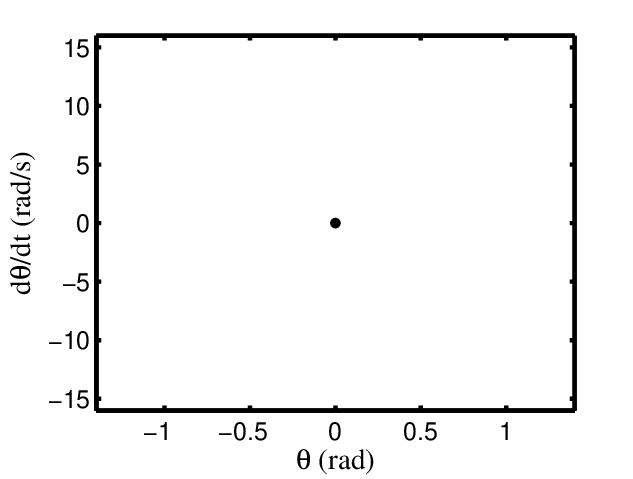}\\
\hspace*{1.0cm} a) \hspace*{4.0cm} b) \hspace*{4.0cm} c) \hfill\\
\caption{\label{FIGURE20}\footnotesize {Different phase portraits of the system. a) Obtained numerically using the first model (\ref{EQUAT6}), b) obtained experimentally, and c) obtained numerically using the second model (\ref{EQUAT6A}), for: $I_{m1}=I_{m2}=1$ A, $f=3.5$ Hz, $\alpha=\frac{\pi}{6}$ rad and $\varphi=\pi$ rad. Black and blue curves are plotted for: $\theta(0)=0.57$ rad,\; $\dot \theta \left( 0 \right)=8.14$ rad/s and $\theta(0)=0.57$ rad,\; $\dot \theta \left( 0 \right)=-8.14$ rad/s, respectively. Magenta and green curves are plotted for: $\theta(0)=-0.57$ rad,\; $\dot \theta \left( 0 \right)=-8.14$ rad/s and $\theta(0)=-0.57$ rad,\; $\dot \theta \left( 0 \right)=8.14$ rad/s, respectively. Finally, red curve is plotted for: $\theta(0)=0.0$ rad,\; $\dot \theta \left( 0 \right)=5.0$ rad/s.}}
\end{center}
\end{figure}

\begin{figure}[h!]
\begin{center}
\includegraphics[width=4.3cm,height=3.8cm]{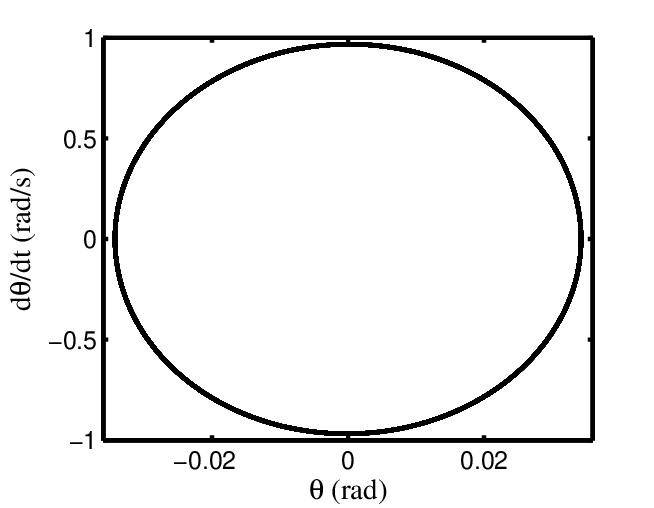}\hspace*{0.0cm}
\includegraphics[width=4.3cm,height=3.8cm]{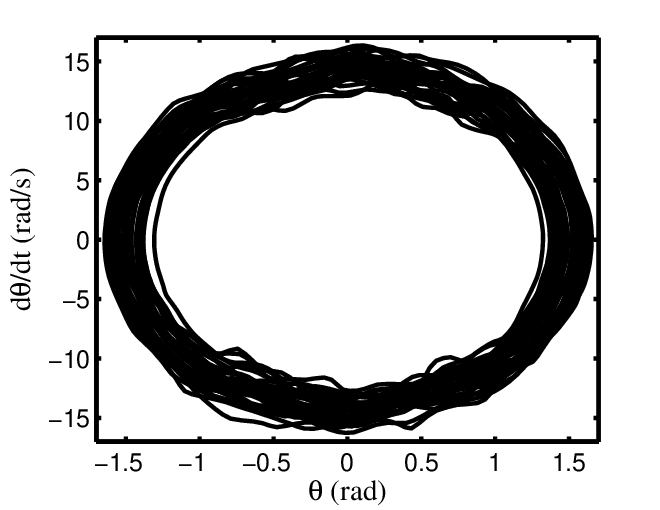}\hspace*{0.0cm}
\includegraphics[width=4.3cm,height=3.8cm]{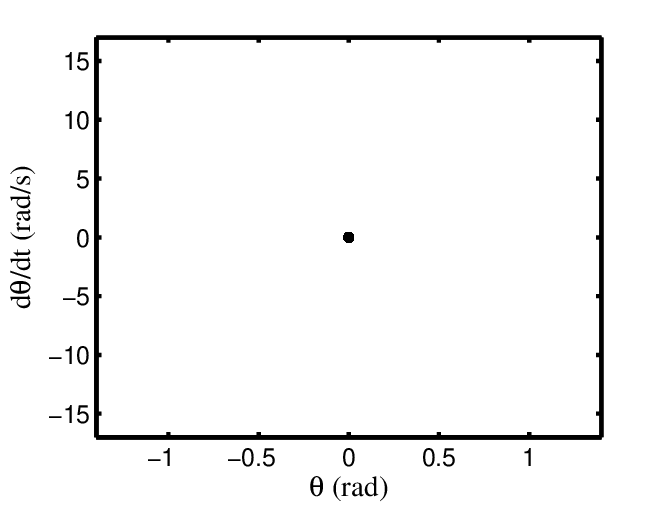}\\
\hspace*{1.0cm} a) \hspace*{4.0cm} b) \hspace*{4.0cm} c) \hfill\\
\caption{\label{FIGURE21}\footnotesize {Different phase portraits of the system. a) Obtained numerically using the first model (\ref{EQUAT6}), b) obtained experimentally, and c) obtained numerically using the second model (\ref{EQUAT6A}), for: $I_{m1}=I_{m2}=500$ mA, $f=4.5$ Hz, $\alpha=\frac{\pi}{6}$ rad and $\varphi=\pi$ rad. The initial conditions for the numerical simulations are $\theta(0)=1.6$ rad,\; $\dot \theta \left( 0 \right)=5.5$ rad/s .}}
\end{center}
\end{figure}

We aim now to visualize the dynamical behavior of the system when the current sources are out of phase $\varphi=\pi$ rad. As we have proceeded above, our results are presented in Figures \ref{FIGURE20} and \ref{FIGURE21}. Figures \ref{FIGURE20}a) and \ref{FIGURE21}a) show the simulation results obtained from the first model while Figures \ref{FIGURE20}c) and \ref{FIGURE21}c) are simulation results obtained using the second model. For comparison, the experimental results are presented in Figures \ref{FIGURE20}b and \ref{FIGURE21}b.

\section{Concluding remarks}

$\quad$ We have analyzed the dynamical states of a magnetic pendulum commanded by two coil magnets disposed symmetrically to the vertical stable position. First considering the coil is powered by constant current, we showed that the system presents different shapes for the potential corresponding to bistable and tristable dynamic states, both of which can be symmetric or asymmetric depending on the sign of the currents and their values. When the currents are sinusoidal functions, two-dimensional bifurcation diagrams (with the frequency and amplitudes of the current as the bifurcation parameters) have been plotted resulting in the system being able to exhibit several dynamical states such as period$-n$ oscillations, chaos, and domains where there is no oscillation although the current is time-dependent.  Plotting the variety of bifurcation diagrams versus the frequency under different initial conditions, it appears that the system can also present multistability with the coexistence of period-1T oscillations and chaos for the same set of parameters. The effects of the value of the separation angle between the coil magnets as well as that of the current being in phase or out of phase have also been analyzed. In the region where the oscillations are of period$-2$, we have employed the harmonic balance method. It has been found that the analytical amplitudes of oscillations have been confirmed by the numerical simulation.

The theoretical part of our paper has been complemented by the experimental investigation. However, one should note that the system is very sensitive to initial conditions leading to multistability so that the correspondence between the theoretical and experimental results is obtained only for specially selected initial conditions.

The results presented in this paper can be used to develop or optimize devices based on magnetic pendulums and aimed at energy recovery. As indicated by earlier works \cite{Kumar2019} in such systems the most energy can be stored in the chaotic regime, therefore further work will be based on the problem of chaos control and destabilization of periodic orbits in magnetic pendulum systems.

\begin{acknowledgements}
The work of the second author K. Polczy\'{n}ski (experiment, numerical calculations, signal processing, and writing) and the last author J. Awrejcewicz (writing, editing, founding) was supported by the  National Science Center, Poland under the grant OPUS 26 No. 2023/51/B/ST8/00021.
\end{acknowledgements}

\bibliography{mybibfile}

\clearpage

\section*{DECLARATIONS}

\section*{Funding}
The work of K. Polczy\'{n}ski and J. Awrejcewicz was supported by the  National Science Center, Poland under the grant OPUS 26 No. 2023/51/B/ST8/00021.

\section*{Conflict of interest}

The authors declare that they have no known competing financial interests or personal relationships that could have appeared to influence the work reported in this paper.

\section*{Availability of data and material}

Not applicable

\section*{Code availability}

Not applicable

\section*{Authors' contributions}

\textbf{Nana B.}: Conceptualization, Methodology, Writing-Original draft preparation.: \textbf{Polczy\'{n}ski K.}: Writing-Reviewing and Editing. \textbf{Woafo P.}: General verification and Editing. \textbf{Awrejcewicz J.}: Supervision.

\section*{Ethics approval}

Informed consent was obtained verbally before submission of this manuscript.

\section*{Consent to participate}

We, Nana B., Polczy\'{n}ski K., Woafo P., and Awrejcewicz J. voluntarily agree to participate in this research study.

\section*{Consent for publication}

We, Nana B., Polczy\'{n}ski K., Woafo P., and Awrejcewicz J. voluntarily agree to submit this research study for possible publication in this Journal (Nonlinear Dynamics).

\end{document}